\documentclass[12pt]{article}
\usepackage{amsmath}
\usepackage{graphicx}
\usepackage{natbib}
\usepackage{url} 

\newcommand{\blind}{0}

\usepackage{mathptmx}
\usepackage{mathtools}
\usepackage{amssymb}
\usepackage{caption}

\usepackage{subcaption}

\usepackage{subfig}
\usepackage{tikz} 
\usepackage{algorithm}
\usepackage[noend]{algpseudocode}

\usepackage{xcolor}

\usepackage{verbatim}
\usepackage{soul}
\usepackage{mycommands}

\sethlcolor{yellow}
\newcommand{\hlc}[2][yellow]{ {\sethlcolor{#1} \hl{#2} \sethlcolor{yellow}} }
\usepackage{array}

\usepackage{rotating}

\renewcommand{\baselinestretch}{1.0}
\DeclareMathAlphabet{\mathbfit}{OML}{cmm}{b}{it}

\begin{document}

\def\spacingset#1{\renewcommand{\baselinestretch}%
{#1}\small\normalsize} \spacingset{1}


\if0\blind
{
  \title{\bf A dimension reduction approach to edge weight estimation for use in spatial models}
  \author{Michael Christensen\hspace{.2cm}\\
    Department of Statistics, Brigham Young University\\
    and \\
    Jo Eidsvik\\
    Department of Mathematical Sciences, NTNU }
  \maketitle
} \fi

\if1\blind
{
  \bigskip
  \bigskip
  \bigskip
  \begin{center}
    {\LARGE\bf Title}
\end{center}
  \medskip
} \fi

\bigskip
\begin{abstract}
Models for areal data are traditionally defined using the neighborhood structure of the regions on which data are observed. The unweighted adjacency matrix of a graph is commonly used to characterize the relationships between locations, resulting in the implicit assumption that all pairs of neighboring regions interact similarly, an assumption which may not be true in practice. It has been shown that more complex spatial relationships between graph nodes may be represented when edge weights are allowed to vary. \citet{christensen2024} introduced a covariance model for data observed on graphs which is more flexible than traditional alternatives, parameterizing covariance as a function of an unknown edge weights matrix. One potential issue with their approach is that each edge weight is treated as a unique parameter, resulting in increasingly challenging parameter estimation as graph size increases. Within this article we propose a framework for estimating edge weight matrices, reducing their effective dimension via a basis function representation. By further leveraging fast matrix calculations for derivative expressions, we enhance the performance and flexibility of covariance models parameterized by such matrices, and demonstrate the utility of our method in a series of illustrations, simulations and data examples.
\end{abstract}

\noindent%
\vfill

\newpage
\spacingset{1.75} 
\section{Introduction}






Studies involving areal, or region indexed data are common in many settings including ecology \citep[e.g.][]{hanks2013}, economics \citep[e.g.][]{arbia2012}, public health \citep[e.g.][]{jin2005} and sociology \citep[e.g.][]{garner1991}. Models for spatially indexed data are typically defined such that observations close to one another in space exhibit higher correlation than observations which are far apart \cite{stein2004approximating}. In contrast to spatial models for point-indexed data, which tend to characterize between-location dependence as a function of the geographic distances between observations, models for areal data generally define spatial dependence in terms of neighborhood or adjacency structure \citep{cressie1993}. 

The spatial structure of areal data may be represented using a graph, with each spatial region as a graph node and with edges existing between each pair of nodes corresponding to physically adjacent regions. First introduced by \citet{besag1974}, the most commonly used spatial model for areal data is the conditional autoregressive (CAR) model. Given vertex set $V$ and edge set $E$, we define the graph $G = (V,E)$ with $p = |V|$ nodes. We observe the random vector $\by$, where $y_1,...,y_p$ are observations at each node of $G$. A CAR model for $\by$ defines a Markov random field where the conditional distribution of each element of $\by$ is determined by the values of $\by$ at adjacent nodes:

\begin{equation}\label{car}
    y_i|\by_{-i} \sim N\left( \sum_{j\sim i} \alpha w_{ij} y_j,\sigma^2\right),
\end{equation}
where $j\sim i$ indicates that node $j$ is adjacent to node $i$, $w_{ij}$ characterizes the strength of the relationship between nodes $i$ and $j$, $\sigma^2 > 0$ is a variance parameter, and $\alpha$ controls the degree of spatial correlation. Equation \ref{car} results in a marginal distribution for $\by$ of
\begin{equation}\label{carmarg}
    \by \sim N_p(\bzero, \sigma^2 (\bI_p - \alpha \bW)^{-1}),
\end{equation}
where $\bW$ is a $p \times p$ adjacency or weights matrix defined such that $w_{ij} > 0$ if $j\sim i$ and $w_{ij} = 0$ otherwise. In order to produce a valid covariance matrix, $\alpha$ must be bounded between the largest and smallest eigenvalue of $\bW$ \citep{verhoef2018a}. An alternative form of the CAR model is
\begin{equation}\label{carmarg2}
    \by \sim N_p(\bzero, \sigma^2 (\text{diag}(\bW \bone_p) - \kappa\bW)^{-1}),
\end{equation}
where $\text{diag}(\cdot)$ returns a diagonal matrix with entries equal to the input vector. This model is valid for all $|\kappa| < 1$. The intrinsic conditional autoregressive (ICAR) model is a special case of Equation \ref{carmarg2}  when $\kappa = 1$. This results in an improper distribution, but it may be reconfigured as a $p-1$ dimensional distribution using the constraint $\sum_{i=1}^p y_i = 0$ \citep{besag1995}.

While many variants of the CAR model and other similarly defined spatial autoregressive models have been developed and utilized, \citet{verhoef2018a} note that in practice almost all such models have been implemented using the unweighted adjacency matrix of $G$, with $w_{ij} = 1$ if $i \sim j$, and $w_{ij}=0$ otherwise. The CAR model is frequently used to characterize spatial random effects or an ``error process" within a larger generalized linear model \citep{bym1991}; in many such contexts, interest in the covariance structure of the model has been secondary to inference on the fixed effects governing the model's mean structure. However, the lack of a more complex weighting scheme results in a pattern of spatial dependence where the partial correlations between all pairs of adjacent regions are {near-identical} regardless of location within the overall spatial domain. Such models may be thought of as discrete space equivalents to stationary and isotropic covariance models for point indexed data in continuous space, properties which are frequently inappropriate for real world data \citep{guttorp1994}.
Despite likely under-utilization in practice, more complex weighting schemes for CAR models have been shown to be {useful}, as in  {\citet{wall2004close} who examines the implied correlation structures in various models for areal data}, and \citet{hanks2013} and \citet{ejigu2020} who define edge weights using linear combinations of environmental covariates. {Weighting schemes with even greater flexibility are also possible and they potentially enable more realistic interaction structures in the discretely observed spatial processes.}

\citet{christensen2024} introduced a covariance model for areal data exhibiting greater flexibility than the traditional CAR model. We refer to this as the graph deformation (GDEF) model. Given the graph $G = (V,E)$, they define $\Sigma = \text{Cov}(\by)$ as follows:

\begin{equation}\label{prevmodel}
\begin{aligned}
    \Sigma_{ij} &= \sigma^2 \rho_\nu(d_{ij}), \\
    d_{ij} &= \sqrt{(\be_i - \be_j)^\top\{\bL^+\}^2(\be_i-\be_j)}, \\
    \bL &= \text{diag}(\bW \bone_p) - \bW, \\
    \sigma^2 &> 0, \; w_{ij} > 0  \text{ if } i \sim j, \text{ else } w_{ij} = 0, \text{ and } w_{ij} = w_{ji}.
\end{aligned}
\end{equation}
Here $\sigma^2$ is a scale parameter and $\rho_\nu(\cdot)$ is the Mat\'{e}rn correlation function \citep{matern1960} with smoothness parameter $\nu$. The term $d_{ij}$ may be thought of as the latent distance between nodes $i$ and $j$ and is defined as a function of unknown edge weights matrix $\bW$. The matrix $\{\bL^+\}^2 = \bL^+\bL^+$ is the square of the generalized inverse of the graph Laplacian $\bL$, and $\be_i$, $i=1,\ldots,p$ are the standard basis vectors. The edge weights $\{w_{ij}\}_{i\sim j}$ are estimated from the observed data {and may be interpreted as the strength of connectivity between adjacent nodes or regions}. Related models have been suggested for resistance distances in physics and chemistry \citep{klein1993}.

The GDEF model described in Equation \ref{prevmodel} results in a valid covariance matrix for any combination of positive edge weights and the matrix $\bW$ is an identifiable parameter. While \citet{christensen2024} demonstrate the flexibility and interpretive value of the GDEF approach, they acknowledge that the model has limited scalability when applied to large graphs and using their Markov chain Monte Carlo (MCMC) algorithm for estimation.

{A key question for both the implementation of the GDEF model and within the literature concerning more complex weighting schemes for spatial autoregressive models is how to define and estimate the edge weights matrix $\bW$. The individual estimation of each unique edge weight, as in \citet{christensen2024}, proves to be computationally impractical for graphs with more than a few hundred nodes, while also introducing potential identifiability issues for models that have not been carefully specified. On the other hand, the covariance structures resulting from previously proposed edge weighting schemes, such as defining edge weights as linear combinations of covariates, or scaling edge weights to be proportional to distances between region centroids or the length of the border shared by two regions, may not be flexible enough to accurately characterize some spatial processes.} 

{The principle contributions from our work within this article are as follows:}

\begin{itemize}
    \item {We introduce a dimension reduction approach for the specification and estimation of an unknown edge weights matrix that utilizes eigenvectors from the Laplacian matrix of a network's line graph as a basis. This method allows for the specification of CAR and other autoregressive models that are considerably more flexible than existing variants, while significantly improving the practical usability of the GDEF framework.}

    \item {By embedding the basis function representation of $\bW$ within a covariance model rather than directly defining spatial random effects as linear combinations of basis functions as has traditionally been done, we leverage the computational advantages of such approaches while producing a class of models with fewer restrictions on the imposed correlation structure.} 

    \item {By incorporating our framework for edge weight estimation within the GDEF framework, we produce a nonstationary spatial covariance model that may be interpreted through the lens of the spatial deformation approach of \citet{sampson1992} that may be fit with no replicate observations of the spatial process.}

\end{itemize}

In Section 2 of the article we provide a brief overview of necessary background information related to spatial basis functions, graphs, and spectral theory. In Section 3 we present our approach to edge weight estimation and illustrate some of its properties when incorporated into existing methods. In Section 4 we illustrate and evaluate the performance of our approach through a series of simulation studies and a real data example. We conclude the article with a discussion of potential applications and extensions.


\section{Background}
 Given a simple, undirected graph $G = (V,E)$, random variables $y_i$, $i \in V$, and defining $\Sigma_{ij} = \text{Cov}(y_i,y_j)$, we parameterize the covariance matrix $\Sigma$ as a function of $\bW$, a matrix of unknown positive edge weights. We define $\mathcal{W}_G$ to be the space of all edge weight matrices $\bW$ that may be defined given $G$ such that $w_{ij} > 0$ if an edge exists between nodes $i$ and $j$ and $w_{ij}=0$ otherwise. {Note that this means that when we speak of estimating $\bW$, we are referring only to the estimation of the edge weights within a fixed network structure, rather than trying to determine the existence or nonexistence of edges between observed nodes.} As the number of nodes in $G$ increases, so too does the size of $\mathcal{W}_G$. 
 To address this challenge, we here restrict the parameter space to $\bW \in \mathcal{W}^B_G \subset \mathcal{W}_G$, where $\mathcal{W}^B_G$ contains all possible edge weights matrices that can be obtained from a basis function representation of the non-zero edge weights.

\subsection{Basis functions and spatial models}

If $y(\bs)$ is an observation of a random process at location $\bs \in \mathcal{D}$ within some spatial domain ($\mathcal{D}$ is typically a subset of $\mathbb{R}^2$, but could also be the set of spatial regions under the areal data setting), with spatially indexed predictors $\bx(\bs)$, a typical model for $y(\cdot)$ is:

\begin{equation}
    y(\bs) = \bx^\top (\bs)\beta + z(\bs) + \epsilon(\bs),
\end{equation}
where $\beta$ is a vector of coefficients, $\bx(\bs)$ include explanatory variables, $z(\bs)$ is a spatial random effect and $\epsilon(\bs) \overset{iid}{\sim}  N(0,\sigma^2)$ for all $\bs \in \mathcal{D}$. 

{Let $\bz = (z(\bs_1),...,z(\bs_p))$ denote the vector of spatial random effects at some collection of $p$ locations.}
Instead of modeling $\bz \sim N_p(\bzero, \Sigma)$ it is common to define $\bz = \bM \eta$, where $\bM$ is an $p \times k$ matrix of $k \ll p$ basis functions, and $\eta \sim N_k(\bzero,\Phi)$. Reduced-rank statistical models representing spatial random effects using basis functions are widespread. Examples of this include, but are not limited to the kernel convolution approach \citep{higdon1998}, predictive processes \citep{banerjee2008}, spatial splines \citep{sangalli2013}, the stochastic partial differential equation approach \citep{lindgren2011}, and various eigenvector and principle component based methods \citep{higdon2008, hughes2013, lee2022}. Note that the majority of these methods were developed for point indexed data, though the framework of \citet{hughes2013} is designed specifically for areal data. Applying methods designed for one data type to another is possible, but requires assumptions that may not be appropriate in all settings \citep{gelfand2001}. 

Basis function methods tend to be very computationally efficient relative to full-rank Gaussian process models. Notably, some approaches to basis functions specification are also capable of addressing issues such as nonstationarity and anisotropy \citep{schmidt2020}. It has often been observed that methods based on low-rank representations of spatial random effects may result in oversmoothing as well as bias in the estimation of variance components \citep{banerjee2010, stein2014, datta2016}. Additionally, when pre-specifying the design matrix $\bM$ using some set of basis functions, the covariance of $\bu$ is restricted to matrices of the form $\bM \Phi \bM^\top$: in many instances, $\Phi$ is assumed to be a diagonal matrix or multiple of the identity matrix, resulting in strong and potentially inappropriate assumptions about model covariance structure. This may be especially true if analysis goals go beyond spatial prediction and smoothing and include inference regarding model covariance. Rather than modeling spatial random effects for areal data as a linear combination of basis functions, we propose using basis functions in a novel parameterization of edge weights matrix $\bW$ which is then incorporated into a CAR or GDEF model, resulting in a flexible class of full-rank covariance matrices that are less computationally demanding than the implementation used by \citet{christensen2024}.

\subsection{Line graphs and an eigenvector basis}
\begin{figure}
\centering

\begin{tikzpicture}[node distance={30mm}, thick, main/.style = {draw, circle}] 
\node[] at (-.2,0) {$G=$};
\node[] at (8,0) {$L(G)=$};

\node[] at (6.5,0) {$\longrightarrow $};

\node[main] (3) at (3,0) {$v_3$}; 
\node[main] (1) [above left of=3] {$v_1$}; 
\node[main] (2) [above right of=3] {$v_2$}; 
\node[main] (4) [below left of=3] {$v_4$}; 
\node[main] (5) [below right of=3] {$v_5$}; 
\draw (1) -- node[midway, above, sloped, pos=.5] {$e_1$} (2); 
\draw (1) -- node[midway, above, sloped, pos=.5] {$e_2$} (3);
\draw (4) -- node[midway, above, sloped, pos=.5] {$e_4$} (1); 
\draw (2) -- node[midway, above, sloped, pos=.5] {$e_3$} (3); 
\draw (3) -- node[midway, above, sloped, pos=.5] {$e_6$} (4);
\draw (2) -- node[midway, above, sloped, pos=.5] {$e_5$} (5); 
\draw (5) -- node[midway, above, sloped, pos=.5] {$e_7$} (3);
\draw (5) -- node[midway, below, sloped, pos=.5] {$e_8$} (4); 

\node[main] (11) at (12,2.7) {$e_1$}; 
\node[main] (12) at (11,1) {$e_2$}; 
\node[main] (13) at (13,1) {$e_3$}; 
\node[main] (14) at (9.3,0) {$e_4$}; 
\node[main] (15) at (14.7,0) {$e_5$}; 
\node[main] (16) at (11,-1) {$e_6$}; 
\node[main] (17) at (13,-1) {$e_7$}; 
\node[main] (18) at (12,-2.7) {$e_8$}; 

\draw (11) -- node[midway, above, sloped, pos=.5] {} (12); 
\draw (11) -- node[midway, above, sloped, pos=.5] {} (13);
\draw (11) -- node[midway, above, sloped, pos=.5] {} (14); 
\draw (11) -- node[midway, above, sloped, pos=.5] {} (15); 
\draw (12) -- node[midway, above, sloped, pos=.5] {} (13);
\draw (12) -- node[midway, above, sloped, pos=.5] {} (14);
\draw (12) -- node[midway, above, sloped, pos=.5] {} (16);
\draw (12) -- node[midway, above, sloped, pos=.5] {} (17);
\draw (13) -- node[midway, above, sloped, pos=.5] {} (15);
\draw (13) -- node[midway, above, sloped, pos=.5] {} (16);
\draw (13) -- node[midway, above, sloped, pos=.5] {} (17);
\draw (14) -- node[midway, above, sloped, pos=.5] {} (16);
\draw (14) -- node[midway, above, sloped, pos=.5] {} (18);
\draw (15) -- node[midway, above, sloped, pos=.5] {} (17);
\draw (15) -- node[midway, above, sloped, pos=.5] {} (18);
\draw (16) -- node[midway, above, sloped, pos=.5] {} (17);
\draw (16) -- node[midway, above, sloped, pos=.5] {} (18);
\draw (17) -- node[midway, above, sloped, pos=.5] {} (18);
\end{tikzpicture}
\caption{The graph $G$ has corresponding line graph $L(G)$. The five nodes are $v_1,\ldots,v_5$. The eight edges are $e_1,\ldots,e_8$.}\label{linegraph}

\end{figure}
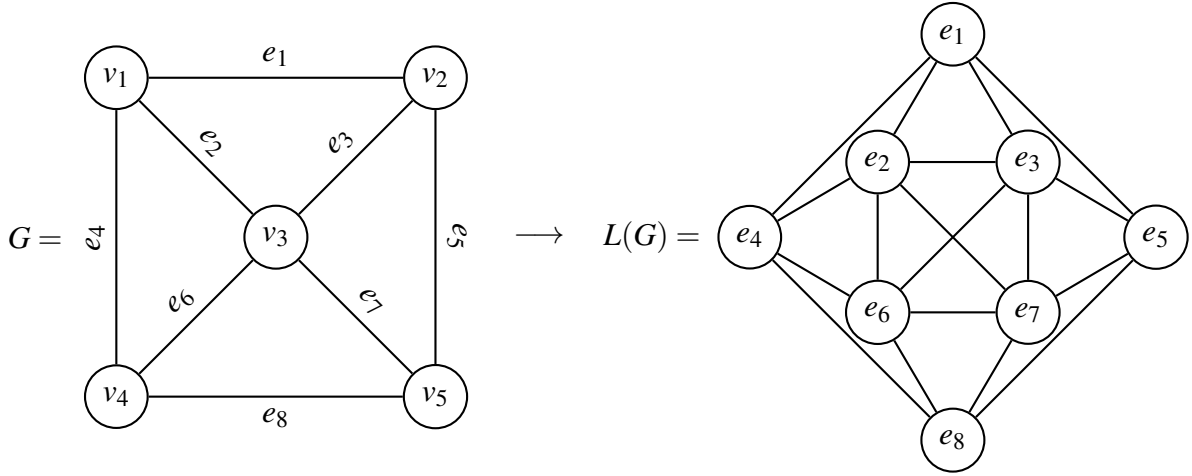

For a graph $G$, one may define its line graph $L(G)$ as a graph with $q$ nodes corresponding to each of the edges of $G$, and edges between each pair of nodes corresponding to coincident edges in the original graph $G$. (Edges are considered coincident if they share a common node). Figure \ref{linegraph} illustrates the relationship between a graph $G$ and its line graph $L(G)$ for a graph containing $p=5$ nodes and $q=8$ edges.

Let $\bA^L$ be the $q \times q$ adjacency matrix of $L(G)$, where $a^L_{ij} = 1$ if nodes $i$ and $j$ of $L(G)$ are adjacent and $a^L_{ij} = 0$ otherwise. Let $\bL^L = \text{diag}(\bA^L \bone_q) - \bA^L$ be the Laplacian matrix of $L(G)$. If $\bV \bLambda \bV^\top$ is the eigendecomposition of $\bL^L$, the columns vectors of $\bV$ form an orthonormal basis for $\mathbb{R}^q$. The eigenvalues of Laplacian matrices are non-negative and generally ordered from smallest to largest, with smallest non-zero eigenvalues corresponding to the most ``important" eigenvectors. Laplacian matrices for connected graphs are positive-semidefinite and have exactly one eigenvalue (denoted $\lambda_1$) equal to 0 with corresponding eigenvector $\bv_1 \propto \bone$. Despite some referring to $\bv_1$ as the ``trivial" eigenvector, it has interpretive value as an intercept column within an eigenvector basis. Considerable work has been done in the field of spectral graph theory investigating the properties of the eigenvalues and eigenvectors of Laplacian matrices \citep{spielman2012}. 

{Laplacian matrices have many applications in various types of data analysis. Within the field of signal processing for data on undirected graphs, the matrix $\bV^{-1}$ defines the graph Fourier transform, and is used to decompose signals in terms of the eigenvectors of the Laplacian \citep{sandryhaila2013discrete, ortega2018graph}. In the field of spatial statistics, the eigenvectors of graph Laplacians have been used as a basis to define a smoothly varying processes over networks, as in \citet{upton2024modeling} who model urban crime on a street network by estimating node-level regression coefficients in such a manner. One of the most novel contributions of our work in this article is the utilization of the eigenvectors of the Laplacian matrix of a network's line graph to define a process over its edges rather than its nodes. Similar methods were applied recently in \citet{josephs2025hypergraph}, who estimated the efficacy of different NBA lineups by utilizing a hypergraph representation of all players and player groupings, though the nature of the networks they analyzed and the random variables modeled differ considerably from those in focus here.}

\begin{figure}[ht]
    \centering
    \includegraphics[width = .75\textwidth]{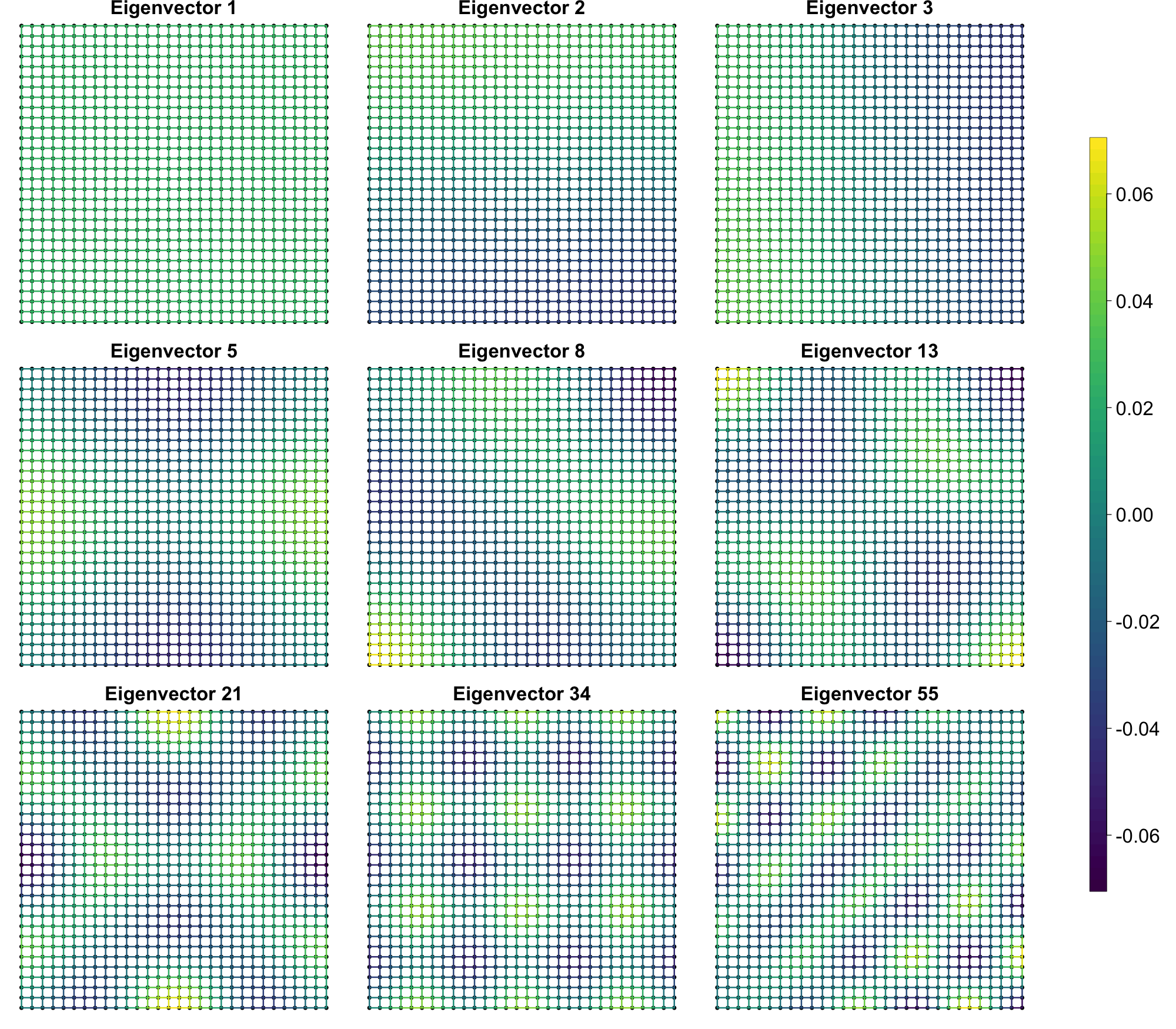}
    \caption{A selection of LGL eigenvectors for the $30 \times 30$ lattice grid.}
    \label{gridbasis}
\end{figure}

\begin{figure}[ht]
    \centering
    \includegraphics[width = .9\textwidth]{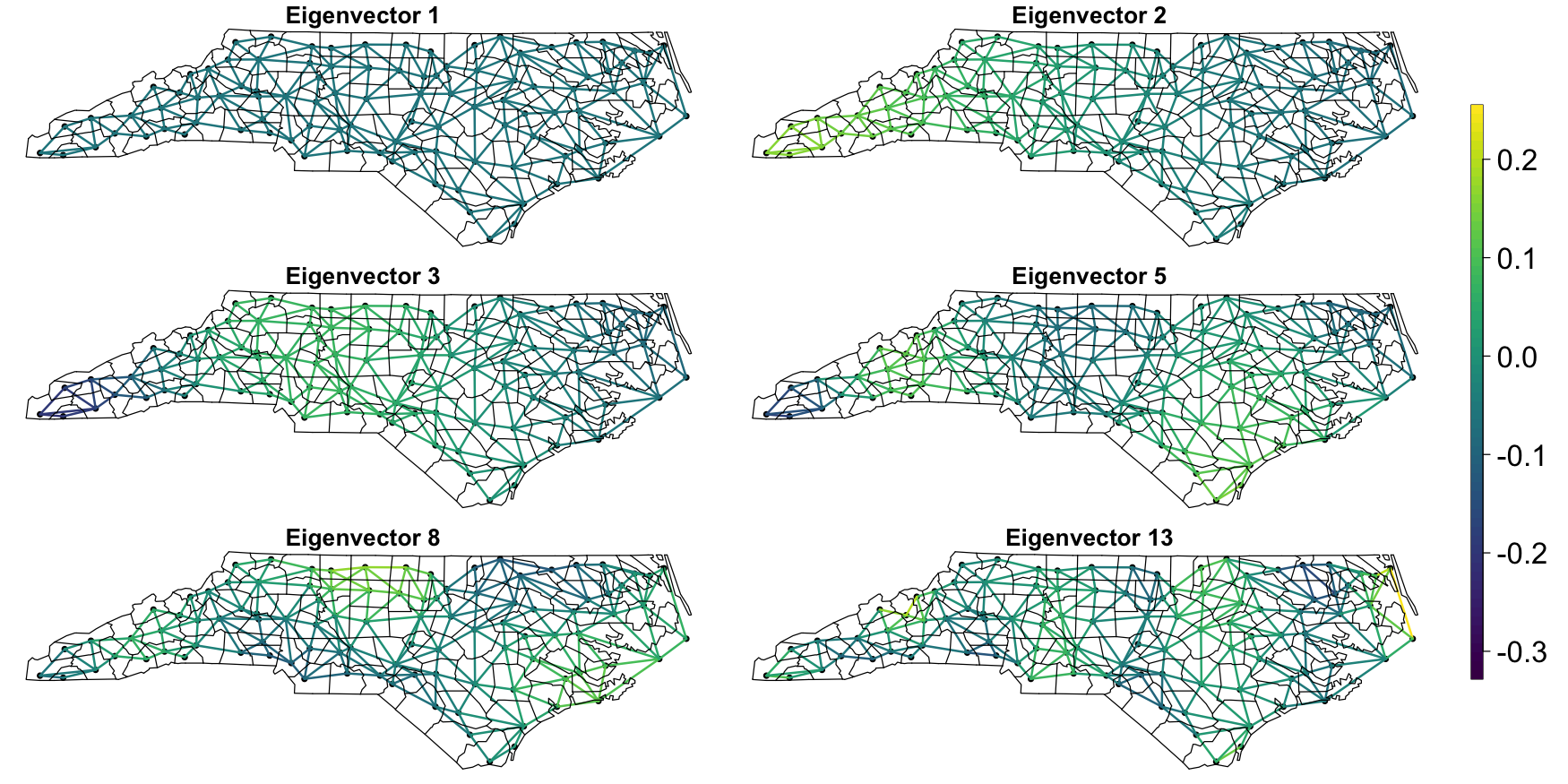}
    \caption{A selection of LGL eigenvectors for the graph of the 100 counties of North Carolina.}
    \label{NCbasis}
\end{figure}

{In order to understand how the eigendecomposition of the line graph Laplacian (LGL) provides a useful basis for modeling the edge weights within a network, we visualize the the original graph with edges weighted by the the corresponding elements of the LGL eigenvectors}. Figure \ref{gridbasis} depicts a selection of LGL eigenvectors for a $30 \times 30$ lattice graph, while Figure \ref{NCbasis} shows that of the 100 counties of North Carolina. Note the harmonic behavior of these eigenvectors, with later eigenvectors exhibiting higher frequency oscillations between positively and negatively valued regions of the graph. This makes selection of the first $k$ eigenvectors (corresponding to the $k$ smallest eigenvalues) a natural choice of basis, with the ``most important" initial eigenvectors capturing larger-scale trends in edge weight behavior. 

\section{Method}

Equations \ref{carmarg2} and \ref{prevmodel} describe two models for areal data, each parameterized by unknown edge weights matrix $\bW$. Let $\bw$ be the $q$-length vector containing the unknown weights associated with the edges of graph $G$. The $p \times p$ edge weights matrix $\bW$ may be constructed from $\bw$ and vice versa. If $\bV \bLambda \bV^\top = \bL^L$ is the eigendecomposition of the LGL, let $\bV^k$ be the $q \times k$ matrix containing the $k$ eigenvectors of $\bL^L$ associated with its $k$ smallest eigenvalues. We can now represent $\bw$ as follows:

\begin{equation}\label{edgebasis}
\begin{aligned}
    \bw &= \text{exp}(\bV^k \eta), \; \eta \in \mathbb{R}^k, \\
\end{aligned}
\end{equation}
where $\eta$ is a $k$-length vector of basis coefficients that fully characterizes the $p \times p$ edge weights matrix $\bW(\eta)$.
For all $\eta \in \mathbb{R}^k$, (and $\sigma^2 > 0$ and $|\kappa| < 1$) the CAR covariance matrix $\Sigma=\sigma^2 (\text{diag}(\bW(\eta) \bone_p) - \kappa\bW(\eta))^{-1}$ and the GDEF covariance matrix defined in Equation \ref{prevmodel} are positive definite matrices. 

{A discussion and comparison of the properties of the CAR and GDEF modeling approaches 
is provided in the supplemental material. In general we find that CAR model is most naturally suited for applications in which the primary focus is the estimation of a dataset's mean structure; in such settings the CAR framework provides an efficient way to account for spatial random effects that may otherwise obstruct inference on the fixed effects. When inference regarding a dataset's covariance structure is an important modeling consideration, many authors have noted that the CAR covariance (even when utilizing weighting schemes) has properties that are undesirable or restrictive in many scenarios. See for example \citet{wall2004close,besag2005,paciorek2013}. While we find that incorporating the edge weighting scheme of Equation \ref{edgebasis} into CAR models improves their flexibility, the primary focus within the remainder of this article will be on the use of the LGL eigenvector basis within the GDEF model.}

\subsection{Visualizations of method and interpretation of basis coefficients}

As discussed in \citet{christensen2024} regarding the GDEF model, each possible edge weights matrix $\bW$ corresponds to an embedding of the graph in high-dimensional Euclidean space which is unique up to isometry. The distances between nodes in this embedding may be thought of as the ``intrinsic" distances between regions. The GDEF model may be viewed as the graph equivalent of the spatial deformation approach for modeling nonstationary data defined by \citet{sampson1992}, where the deformation function is obtained by estimating the edge weights. This interpretation leads us to say that nodes in regions containing high valued edges are ``closer" to each other than regions with low valued edges. While the CAR model does not directly admit a distance based interpretation of $\bW$, it can be said of both the CAR and GDEF approaches that edges with large weights correspond to greater connectivity between incident nodes.

Let $\bW(\eta)$ be the $p \times p$ matrix of edge weights parameterized by $\eta \in \mathbb{R}^k$. Equation \ref{edgebasis} states that the log of the $q$-length vector containing the unique edge weights in $\bW(\eta)$ is a linear combination of a collection of $k$ basis functions of length $q$, with $\bV^k$ defined as the $k$ eigenvectors of the LGL corresponding to its $k$ smallest eigenvalues. We wish to provide some insight into the interpretation of the coefficient vector $\eta = (\eta_1,...,\eta_k)'$ and illustrate some of the covariance patterns this construction can produce.

\begin{figure}
    \centering
    \includegraphics[width = .95\textwidth]{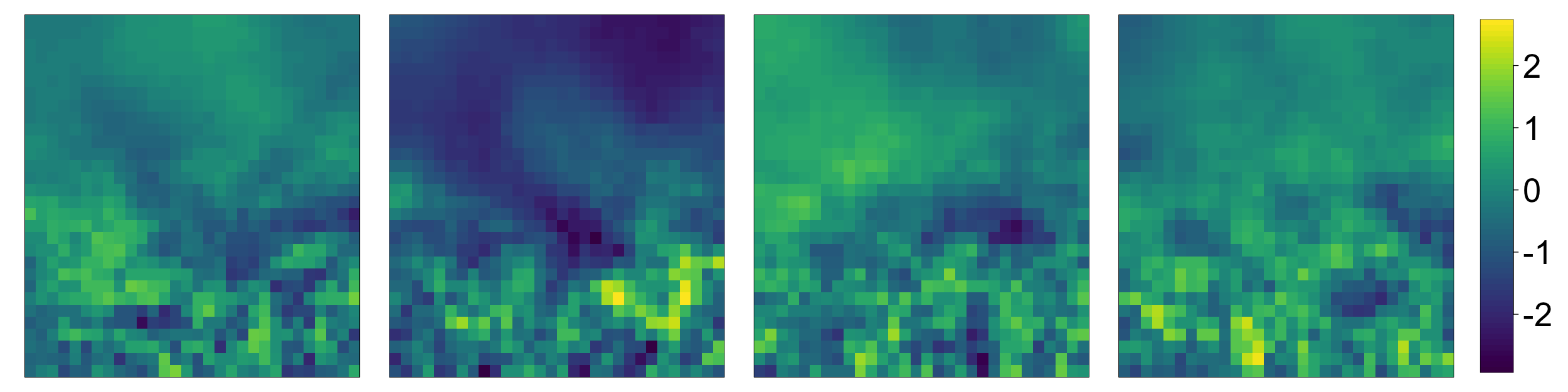}
    \caption{Samples from $N(\bzero,\Sigma)$ over a $30 \times 30$ grid using two basis functions with $\eta_1=\eta_2=50$. Note how spatial correlation decays at different rates across the spatial domain.}
    \label{gdef_samps}
\end{figure}

Recall that $\eta_1\bv^k_1$ is constant and may be viewed as the intercept for the log edge weights. Because $\text{exp}(\bV^k\eta) = \text{exp}(\eta_1\bv^k_1)\odot\text{exp}(\bV^k_{-1}\eta_{-1})$, the vector $\text{exp}(\eta_1\bv^k_1)$ may be viewed as a scaling factor for the edge weights matrix $\bW(\eta_{-1})$ which is defined using the other $k-1$ basis functions. In the GDEF model, this scaling factor is directly related to the range of spatial correlation, while in the CAR model, the scaling factor is redundant when $\sigma^2$ is estimated. As such, we recommend fixing $\eta_1=0$ when using the basis function approach to characterizing the edge weights matrix for CAR models. Subsequent coefficients in $\eta$ may be interpreted in context of the eigenvectors with which they are associated. As depicted in Figure \ref{gridbasis}, the second LGL eigenvector of a $30 \times 30$ grid characterizes a contrast between the top and bottom of the spatial domain. Figure \ref{gdef_samps} depicts samples {from a GDEF covariance model defined using only the first two LGL eigenvectors and smoothness parameter $\nu = 3/2$. Across samples it can be seen how observations in the top of the spatial domain are highly correlated, while the effective range of spatial correlation is much smaller on the opposite side. Spatial covariance patterns such as this do appear in real world settings; for example \citet{paciorek2006} present a data set for precipitation in Colorado in which observations east of the Rocky mountains are highly correlated over long distances, whereas spatial correlations between observations in the mountainous western half of the state decay rapidly over short distances.}

\begin{figure}[h]
    \centering
    \includegraphics[width = .8\textwidth]{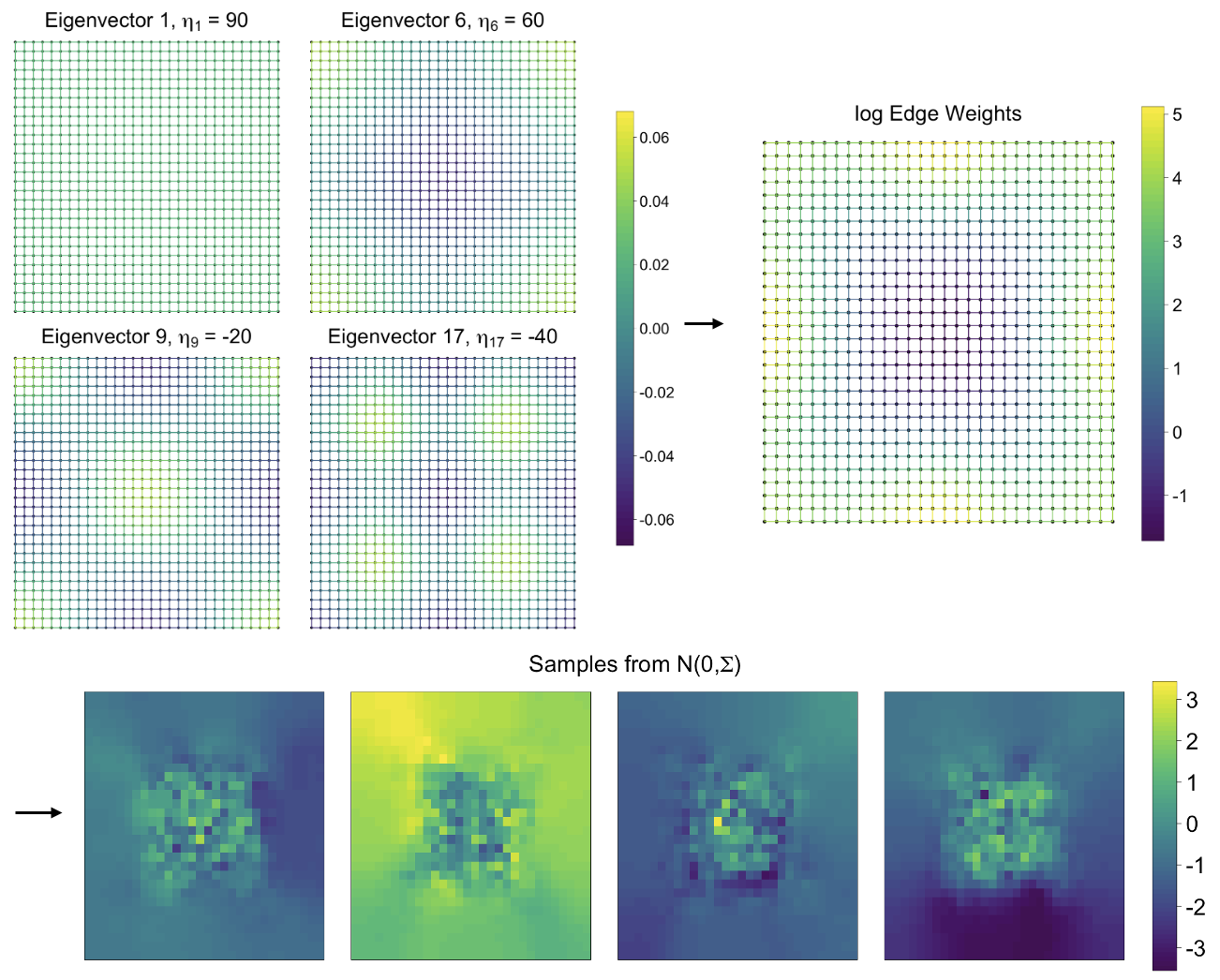}
    \caption{Illustration of how limited number of basis functions may combine to produce complex covariance structures. $\Sigma$ defined using GDEF covariance with $\nu=3/2$.}
    \label{example}
\end{figure}

Different linear combinations of basis functions can produce a wide array of covariance structures. 
Figure \ref{example} depicts a linear combination of four eigenvectors that results in covariance where observations at the borders of the spatial domain are highly correlated, while observations in the interior are not. In effect locations at opposite corners of the grid are ``closer" together than locations that are both near the center. The more basis functions are included, the more flexible the covariance, though there are generally diminishing returns as the number of eigenvectors used increases due to the higher spatial frequencies exhibited by later eigenvectors. Omitting these high frequency eigenvectors essentially assumes that the edge weights themselves exhibit some degree of spatial smoothness. The question of how to select the number of basis functions is explored later in Section 4, and a discussion regarding the potential inclusion of environmental covariates within our model for edge weights is provided in the supplemental material.

\subsection{Parameter estimation and model fitting}

{Much of the spatial deformation literature assumes settings in which there are replicate observations of the spatial process whose covariance is being estimated. Commentary on this assumption is provided in the supplemental material. Should we have}
replicate observations, we record them as $\by_t \in \mathbb{R}^p$ for $t \in 1,...,n$. These observations may be stored in the $n \times p$ data matrix $\bY$, where $y_{ti}$ is the $t$th observation at the $i$th node of graph $G$. For now, we assume that $\by_t \overset{iid}{\sim} N_p(\bzero,\Sigma)$, with $\Sigma$ parameterized by $\theta = (\sigma^2,\eta)$ for the GDEF model and $\theta = (\sigma^2,\eta,\kappa)$ for the CAR model. \citet{christensen2024} only considered settings with multiple observations of the spatial process. This was necessary due to each edge weight being a unique parameter and the fact $q > p$ in almost all settings. Importantly, the dimension reduction of the basis function approach to estimation of $\bW$ allows us to model the covariance of spatial processes with only a single realization ($n=1$).

The maximum likelihood estimator (MLE) for $\theta$ may be obtained using {Fisher scoring} which requires computation of the first and second derivatives of the log likelihood $\ell(\theta;\bY)=-\frac{n}{2} \log |\bSigma(\btheta)|-\frac{1}{2}\sum_{t=1}^n \by^T_t \bSigma^{-1}(\btheta) \by_t$ with respect to $\theta$. Given data $\bY$ and an initial set of parameter values $\theta^0$ we can maximize the likelihood by iteratively calculating
\begin{equation}\label{NR}
    \theta^{l}=\theta^{l-1}- \gamma \; \mathbb{E}\left[\bH_\ell(\theta^{l-1}) \right]^{-1} \left( \nabla_\ell(\theta^{l-1})\right), \hspace{5mm} l=1,2,\ldots,
\end{equation}
until $\theta^{l}$ converges to the MLE. Here $\gamma$ is a scaling parameter between $0$ and $1$. Smaller $\gamma$ provides more robust optimization, while $\gamma=1$ will generally be faster. {Identifying the impact of individual basis function coefficients on the likelihood is fairly challenging, but} details on the efficient matrix forms of $\nabla_\ell(\theta) = \frac{\partial \ell(\theta;\bY)}{\partial \theta}$ and $\bH_\ell(\theta) =  \frac{\partial^2 \ell(\theta;\bY)}{\partial \theta^2}$ are provided in the Supplemental material. 

Uncertainty quantification may be conducted using the approximation
\begin{equation}\label{MLEapprox}
    \hat{\theta} \sim N\left(\theta,-\left[\bH_\ell(\hat{\theta})\right]^{-1}\right).
\end{equation}
Note when employing the algorithm in Equation \ref{NR} that we use the expectation of the Hessian matrix $\bH_\ell(\theta)$, which is less computationally intensive to obtain than the Hessian matrix itself. We only need to compute $\bH_\ell(\theta)$ once after obtaining the MLE in order to use the approximation in Equation \ref{MLEapprox}.


\subsection{Prediction}

{The correlation structures of most models for areal data, including the GDEF and especially the CAR models, are sensitive to the structure of the graph on which they are defined \citep{wall2004close,christensen2024}. As such, prediction for new or unobserved locations has to be approached more carefully than when using models for point-indexed data to ensure validity.}

{The log likelihood of $\by_t$ and its derivatives with respect to $\theta$ are still defined when $\by_t$ is not fully observed at all nodes within a graph. If one wishes to make predictions for unobserved nodes or new locations, one simply needs to ensure that the graph $G$ used to fit the model is specified such that its nodes and edges are defined for all relevant locations, both where data are observed and where one wishes to predict. Because the LGL basis functions used to characterize the edge weights matrix $\bW$ are defined over the full network rather than locally, edge structure in unobserved regions of the graph is still estimable. After obtaining estimates for all parameters needed to define $\Sigma(\theta)$, it is straightforward to use the conditional distribution of $\by_\text{predicted}|\by_\text{observed}$ to obtain predictions and prediction intervals.}

\section{Illustrations}

In this section we evaluate the performance of the GDEF model using the LGL eigenvector basis in both simulated and real-world data settings. {We begin with two simulation studies. In the first, we consider how to select the number of basis functions when implementing our method. The second simulation assesses model performance under misspecification, evaluating the Kullback-Leibler divergence between the true data generating distribution and the estimated distribution. The Supplemental material contains an additional simulation study  demonstrating coverage rates of the approximate confidence intervals obtained using Equation \ref{MLEapprox}, as well as additional details on the two simulations presented here.}

\subsection{Simulation 1: Number of basis functions}
{
We defined $10 \times 10$ and $20 \times 20$ lattices obtaining LGL eigenvectors for each. Instead of simulating edge weights directly from Equation \ref{edgebasis}, we generate an edge weights matrix $\bW$ from an ICAR model using the line graph adjacency matrix to define the associations between edges. Figure 3 of the supplemental material for this article depicts one such sampled set of edge weights over a $20 \times 20$ lattice. The spatial dependence in the simulated edge weights themselves means that $\bW$ can be well approximated using our method. Larger numbers of basis functions will improve this approximation, but at the cost of model complexity and potential computational instability. The simulated $\bW$ was used to define a covariance matrix according to Equation \ref{prevmodel} with $\nu = 3/2$ and $\sigma^2=1$. Independent samples were generated from $N_p(\bzero,\Sigma)$ and stored in $n \times p$ matrix $\bY$ for $n \in \{1,10,50\}$. For each simulated $\bY$ we fit a GDEF model using Mat\'{e}rn covariance with $\nu=3/2$ using $k \in \{10,20,...,100\}$ basis functions and evaluated AIC and BIC for each of the 10 model fits. The simulation was repeated 10 times. Table \ref{sim2} contains information regarding the AIC- and BIC-preferred choices of $k$.}

\begin{table}[ht]
\centering
\setlength\extrarowheight{-6pt}
\begin{tabular}{l|rrrr}
  \hline
 & Mode $k$ (AIC) & Mean $k$ (AIC) & Mode $k$ (BIC) & Mean $k$ (BIC) \\ 
  \hline
$p=100, \; n=1$ & 10 & 12 & 10 & 10 \\ 
  $p=100, \; n=10$ & 40 & 58 & 10 & 11 \\ 
  $p=100, \; n=50$ & 100 & 100 & 40 & 42 \\ 
  $p=400, \; n=1$ & 20 & 37 & 10 & 10 \\ 
  $p=400, \; n=10$ & 100 & 99 & 30 & 39 \\ 
  $p=400, \; n=50$ & 100 & 100 & 100 & 99 \\ 
   \hline
\end{tabular}
\caption{Mode $k$ indicates the most commonly selected number of basis functions, while Mean $k$ indicates the average number of selected basis functions. Larger grids and larger sample sizes result in preference for higher $k$.}\label{sim2}
\end{table}

{Across simulation settings, we see that models with higher $k$ perform better on the larger grid and with bigger sample sizes. 
Using the information in Table \ref{sim2}, a potential starting point is to select $k=\sqrt{np}$ when the model reasonably approximates the true data generating process, as was the case within this simulation. Optimal choice of $k$ will also be contingent on the nature of the spatial process itself, as evidence of more complicated patterns of spatial dependence may indicate that larger numbers of basis functions should be used.} 


\subsection{Simulation 2: Model performance under misspecification}

\begin{figure}
    \centering
    \includegraphics[width = \textwidth]{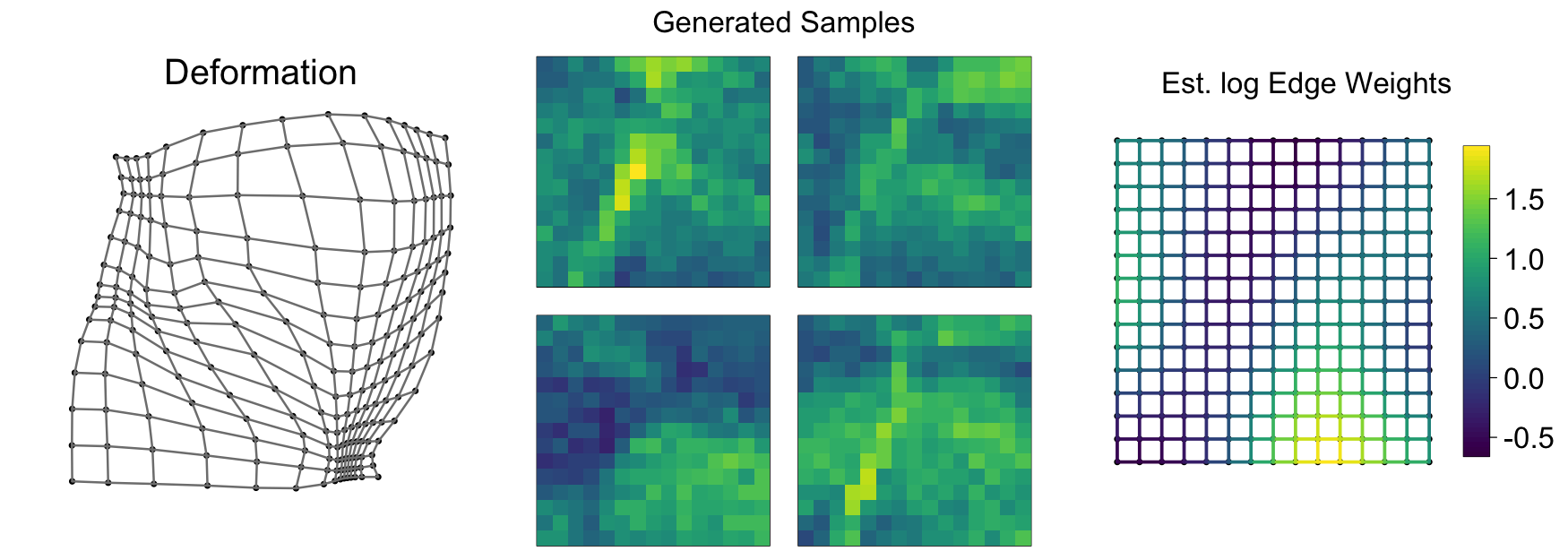}
    \caption{Distances between deformed grid locations define a nonstationary covariance matrix $\bSigma$. Data $\bY$ generated from $N_p(\bzero,\Sigma)$. GDEF model is fit to $\bY$, producing the depicted $\hat{\bW}$ and corresponding $\hat{\Sigma}^{GDEF}$.}
    \label{def_ill}
\end{figure}

We compare the performance of the GDEF model with others in a setting where the GDEF model is misspecified. Data is generated by the nonstationary spatial process obtained by the spatial deformation approach of \citet{sampson1992}. This defines the process covariance
\begin{equation}
    \text{Cov}(z(\bs),z(\bs')) = \sigma^2 \rho(||g(\bs)-g(\bs')||),
\end{equation}
where $g(\cdot)$ is a transformation from the geographic location $\bs$ to its location $g(\bs)$ in a ``deformed" space wherein covariance is stationary and isotropic. 
Data are simulated by first defining a $15 \times 15$ lattice. 
We then generate random deformations of the lattice by contracting or expanding the grid towards or away from randomly sampled locations.
Let $\bD^D$ be the $225 \times 225$ matrix of Euclidean distances between the nodes of the deformed lattice. Applying a Mat\'{e}rn correlation function to $\bD^D$ results in a covariance matrix that is nonstationary with respect to $\bD$, the matrix of Euclidean distances between the original grid point locations. We generate $n$ observations from the following model
\begin{equation}\label{sim3}
\begin{aligned}
    \by_t &= \beta_0\bone + \bz_t + \epsilon_t, \\
    \bz_t &\sim N_p(\bzero,\sigma^2\rho_{5/2}), \\
    \epsilon_t &\sim N_p(\bzero,\tau^2\bI_p),
\end{aligned}
\end{equation}
where $\bz_t$ is a vector of spatial random effects with correlation matrix $\rho_{5/2}$ defined using a Mat\'{e}rn correlation function with smoothness $\nu=5/2$.
We set $\beta_0 = 0$, $\sigma^2=0.9$, $\tau^2 = 0.1$ and a range parameter of $3$. We fit the GDEF and other models to the simulated data $\bY$ as illustrated by Figure \ref{def_ill}, which depicts a deformed lattice, samples of $\by_t$ produced by that deformation, and an estimated edge weights matrix. 

We randomly generated 125 unique grid deformations. For each deformation we sampled $\bY$ as described above for $n \in \{1,2,3,5,10,25,50\}$ and fit Equation \ref{sim3} while defining the covariance of the random effects $\bz_t$ using the following five submodels: a GDEF model with $\nu = 3/2$ and $k = 15$ basis functions, the spatial deformation approach of \citet{sampson1992} (S\&G) as implemented by the \texttt{deform} R package \citep{deformR}, a simple Mat\'{e}rn covariance model with $\nu = 5/2$ and using the Euclidean distances between non-deformed lattice nodes, the method presented in \citet{hughes2013} (H\&H) for modeling random effects in areal data using 50 basis functions, and a standard CAR model using an unweighted adjacency matrix. The spatial deformation approach and the stationary Mat\'{e}rn covariance model treat the data as though it were point-indexed, while all other approaches utilize only the lattice's graphical structure; note however, that the true data generating process exists in continuous space.

Let $\Sigma = \sigma^2\Phi + \tau^2 \bI_p$ be the marginal covariance of $\by_i$ integrated over the random effects $\bz_i$. For each $\bY$ and fitted model, we obtain the MLE $\hat{\Sigma}$ and compute $\text{KL}(N_p(\bzero,\Sigma),N_p(\bzero,\hat{\Sigma}))$, the Kullback-Leibler  (KL) divergence between the true data generating distribution and the distribution characterized by the estimated covariance. As both distributions are mean-zero and Gaussian, KL divergence may be computed using only the covariance matrices $\Sigma$ and $\hat{\Sigma}$:
\begin{equation}
    \text{KL}(\Sigma,\hat{\Sigma}) = \frac{1}{2}\left( \text{log} \frac{|\hat{\Sigma}|}{|\Sigma|} + \text{tr}\left(\hat{\Sigma}^{-1}\Sigma)\right)-p\right).
\end{equation}

Table \ref{KL} depicts the average KL divergence between the true distributions and the distributions estimated by the five models. Note that the true distribution is different for each of the 125 replicates which used a unique deformation. We also consider the frequency with which each model had the smallest KL divergence under the same settings. As can be seen, the GDEF model performs well across sample sizes, {though the simple Mat\'{e}rn and CAR models frequently produce competitive estimates of $\Sigma$ when $n=1$. It is intuitive that simpler models would perform better in contexts with limited data available, and visual inspection 
of simulation output suggests that the simple Mat\'{e}rn and CAR models perform best in comparison to the GDEF approach when the randomly generated deformation exhibits less dramatic patterns of warping. The GDEF model is the best performing in settings with 10 or fewer replicates, though in settings with $n=25$ and $n=50$ it is surpassed on average by the S\&G approach, which under the settings of this simulation study is closest to the true data generating model. The fact that the GDEF model consistently outperforms S\&G for $n \leq 10$ suggests that its internal structure provides some degree of stability and regularization when estimating complex covariance structures using limited numbers of replicates.} 

\begin{table}[ht]
\centering
\setlength\extrarowheight{-6pt}
\begin{tabular}{rlllll}
  \hline
 & GDEF & S\&G & Mat\'{e}rn  5/2 &  H\&H & Std. CAR \\
 \hline
$n=1$ & 42.54 (0.552) & - & 45.22 (0.24) & 70.85 & 43.4 (0.208) \\ 
  $n=2$ & 28.2 (0.928) & 73.6 & 42.35 (0.04) & 69.35 & 42.28 (0.032) \\ 
  $n=3$ & 24.24 (0.984) & 40.29 & 40.87 (0.016) & 75.78 & 41.88 \\ 
  $n=5$ & 21.93 (0.992) & 32.33 (0.008) & 39.75 & 58.58 & 41.67 \\ 
  $n=10$ & 20.6 (0.896) & 26.54 (0.104) & 39.13 & 51.75 & 41.56 \\ 
  $n=25$ & 19.83 (0.376) & 19.09 (0.624) & 38.65 & 50.10 & 41.44 \\ 
  $n=50$ & 19.55 (0.192) & 17.02 (0.808) & 38.36 & 49.7 & 41.39 \\ 
   \hline
\end{tabular}
\caption{Average $\text{KL}(\Sigma,\hat{\Sigma})$ for each model and number of samples. In parentheses is the frequency across simulations that a given model produced the smallest KL divergence for a sample of size $n$. The GDEF model performs well across settings, and especially with limited numbers of replicates.}\label{KL}
\end{table}

Note that the spatial deformation approach (S\&G) cannot be fit for $n=1$. This highlights one of the key advantages of our approach. Much of the literature regarding the spatial deformation framework requires that the spatial process be observed multiple times, and in some implementations requires that $n > p$. However, it is extremely common for spatial analyses to be conducted on datasets with little or no replication. Whether the deformation function is estimated using thin plate splines as in \citet{sampson1992} or a Gaussian process prior as in \citet{schmidt2003}, the inherent flexibility of the framework requires considerable amounts of data (or strong prior assumptions) to avoid over-fitting. As seen here the GDEF model using the basis function representation of $\bW$ as presented in this paper can effectively estimate complicated patterns of spatial dependence for areal data and performs well in a setting where the true spatial process exists in continuous rather than discrete space.

\subsection{Example: Mercer and Hall wheat yield data}
\begin{figure}
    \centering
    \includegraphics[width = .45\textwidth]{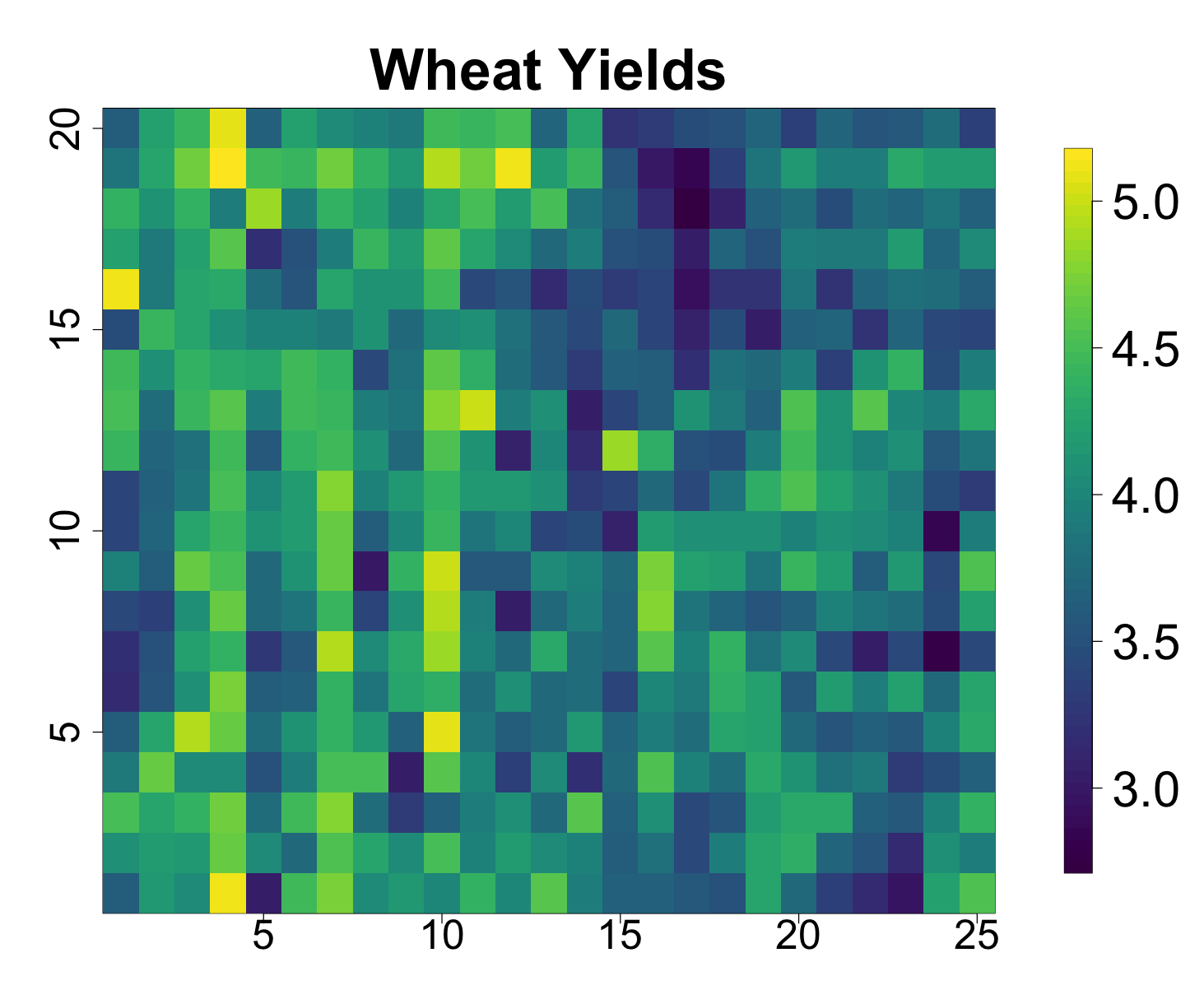}
    \caption{The Mercer Hall wheat yield data. Each cell corresponds to an agricultural plot; response variable is pounds of grain at harvest.}\label{wheatdata}
\end{figure}

We conclude this section with an analysis of the classic wheat yield dataset from \citet{mercer1911} and which is available in the \texttt{spData} R package. The version of the data used in the package was taken from \citet{cressie1993}. \citet{mercer1911} considered a one acre parcel of farmland that was divided into 500 rectangular plots and planted with wheat. Each plot received the same treatment and soil quality was perceived to be essentially uniform at the time of planting. At harvest, the yield of each plot in pounds of grain was recorded. Figure \ref{wheatdata} depicts the yields for each plot, which were arranged in 20 rows each containing 25 plots. Mercer and Hall discuss the variability present in the data and explore the possibility of location-based effects on yield. Noting evidence of spatial correlation, the wheat yield data was analyzed in \citet{whittle1954}, which first introduced the simultaneous autoregressive (SAR) model, and in \citet{besag1974}, which first introduced the CAR model. The authors of both papers found the fit of their models to the wheat yield data unsatisfactory.  Columns (4,7 and 10) in Figure \ref{wheatdata} have larger average yields than others and the within-column correlation appears to be stronger than within-row correlation. By modeling the data using fixed row and column effects \citep{cressie1993} found that a stationary and isotropic kriging model with plot locations given by centroids resulted in satisfactory fit, though it is worth considering whether the inclusion of row and column effects in a model for this data provide interpretive value or merely serve to improve the fit of an otherwise poorly specified model.

\begin{figure}
    \centering
    \includegraphics[width = .9\textwidth]{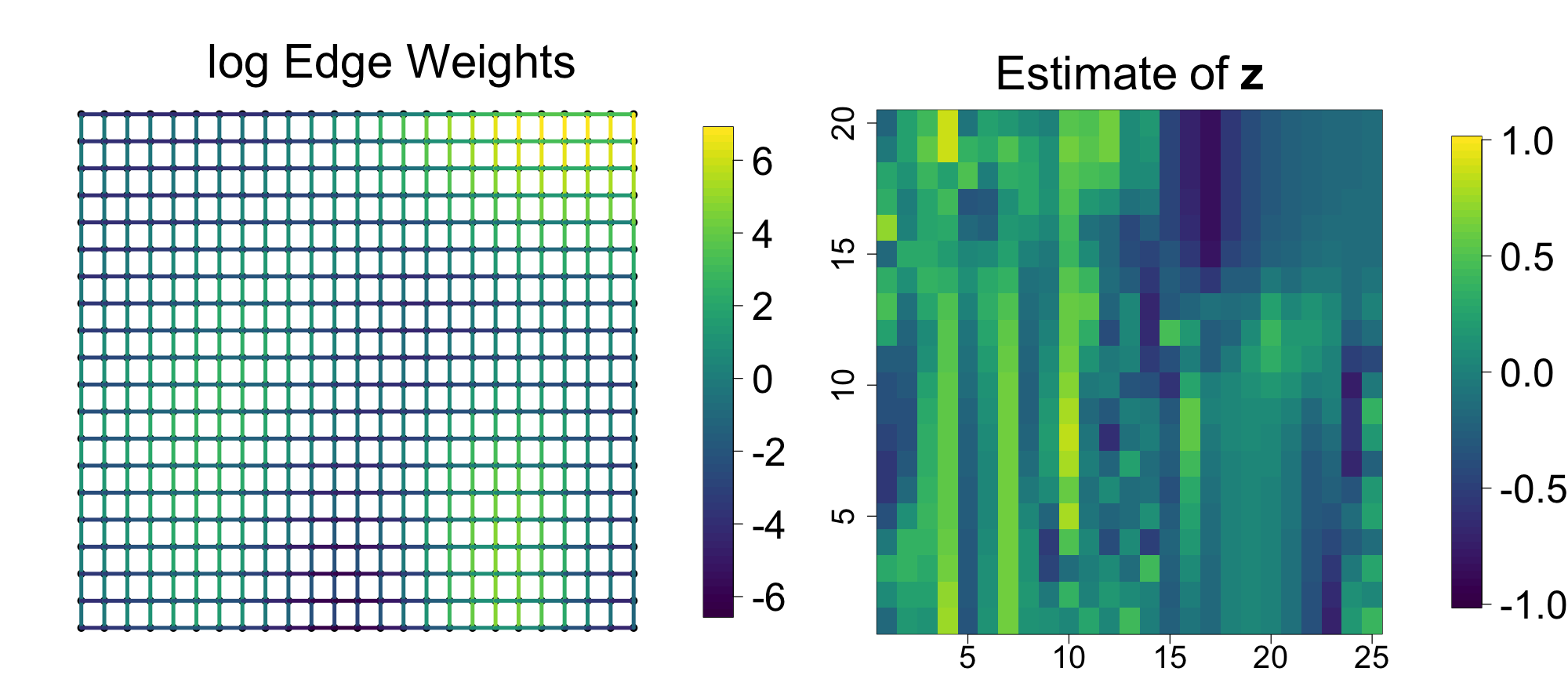}
    \caption{Estimated log edge weights (left) for the Mercer-Hall data which in turn define the covariance structure of $\bz$, the ``de-noised" version of the centered data (right).}\label{wheatfit}
\end{figure}

We propose modeling this data according to Equation \ref{sim3} with the covariance of the random effects defined by a GDEF model with smoothness $\nu=3/2$ and $k=20$ basis functions. To allow for a distinction between within-column and within-row dependency, we split the intercept vector $\bv_1$ of our basis into two vectors, $\bv_1^{row}$ and $\bv_1^{col}$, where $v_{i1}^r = 1$ if edge $i$ connects two nodes in the same row and is 0 otherwise, with $\bv_1^{col}$ defined similarly for columns. We obtain maximum likelihood estimates $\hat{\eta}$, $\hat{\sigma^2}$, $\hat{\tau^2}$ and $\hat{\beta_0}$ which are reported in Table \ref{wheatMLE}. Negative $\hat{\eta_1}^{row}$ and positive $\hat{\eta_1}^{col}$ indicates much stronger within-column than within-row correlations. Rather than reporting the other coefficients in $\hat{\eta}$ which are individually less interpretable, the left half of Figure \ref{wheatfit} depicts log($\hat{\bw}$) over the entire network; there are three regions within the spatial domain with larger edges indicating stronger spatial correlation between nodes in those locations. 

\begin{table}[ht]
\centering
\setlength\extrarowheight{-6pt}
\begin{tabular}{lrr}
  \hline
 & Est. & 95\% CI  \\ 
  \hline
$\eta_1^{row}$ & -2.195 & (-3.551, -0.838) \\ 
  $\eta_1^{col}$ & 1.308 & (0.286, 2.329) \\ 
  $\sigma^2$ & 0.146 & (0.101, 0.191) \\ 
  $\tau^2$ & 0.073 & (0.027, 0.118) \\ 
  $\beta_0$ & 3.949 & (3.871, 4.026) \\
   \hline
\end{tabular}
\caption{MLEs and confidence intervals for model parameters.}\label{wheatMLE}
\end{table}

We can also obtain an estimate of $\bz$, which may be thought of as the smoothed or ``de-noised" version of the centered data. Because $\by$ and $\bz$ are each normally distributed, the conditional distribution of $\bz$ is straightforward to obtain:
\begin{equation}
    \bz|- \sim N_p\left(\left( \Phi^{-1}/\sigma^2+\bI_p/\tau^2\right)^{-1}(\by-\beta_0\bone)/\tau^2, \left( \Phi^{-1}/\sigma^2+\bI_p/\tau^2\right)^{-1}\right).
\end{equation}
We define $\hat{\bz} = \left( \hat{\Phi}^{-1}/\hat{\sigma^2}+\bI_p/\hat{\tau^2}\right)^{-1}(\by-\hat{\beta_0}\bone)/\hat{\tau^2}$ and provide a plot of $\hat{\bz}$ in the right half of Figure \ref{wheatfit}. {Note that regions where the estimate of $\bz$ are smoother correspond to larger estimated edge weights.} Moran's I \citep{moran1950} is a test statistic indicating the presence or absence of spatial autocorrelation. Moran's I for the estimated residuals of our model $\hat{\epsilon}= \by-\hat{\beta_0}\bone -\hat{\bz}$ is 0.022, with a p-value of 0.460, indicating a lack of spatial correlation in $\hat{\epsilon}$. The Shapiro-Wilk normality test \citep{shapiro1965} on the estimated residuals results in a p-value of 0.262 indicating that the normality assumption for $\epsilon$ is not violated.



{
To assess the out-of-sample predictive performance of the GDEF model using the LGL eigenvector basis on this data set in comparison to existing methods, we perform ten-fold spatial cross validation, partitioning the 500 plot locations into 20 five-by-five blocks of plots, then splitting the 20 blocks into ten sets of two, resulting in a total of 50 plots per split. In addition to the GDEF model described above, we fit each of the following models to the test-training splits: a Mat\'{e}rn covariance model with $\nu = 5/2$ and distances computed using plot centroids, the spatial random effects model of \citet{hughes2013} using 50 basis functions, a standard CAR model, and a CAR model with two correlation parameters, one for each of the within-row and within-column dependencies, a modeling consideration similar in nature to the split we made of the intercept column in the basis representation of the GDEF edge weights. For each model and test-training split, we compute the root mean square error (RMSE) and mean absolute error (MAE), and produce 95\% prediction intervals for all elements of the test set. In addition to our comparison of predictive performance, we wish to evaluate the appropriateness of each model for the wheat yield data using residual diagnostics. Each model was fit using the full data set and estimated residuals were obtained in a manner similar to the process described in the previous paragraph for the GDEF model. Moran's I is computed for each model to assess whether the estimated residuals had any spatial dependence and the Shapiro-Wilk test is applied to assess the normality of the residuals. The left side of table \ref{cvresults} contains the cross-validated average RMSEs, MAEs, coverage rates, and prediction interval widths for each model, while the right side contains the p-values associated with the Moran's and Shapiro-Wilk tests.
}
\begin{table}[ht]
\centering
\setlength\extrarowheight{-6pt}
\begin{tabular}{r|rrrr|rr}
  \hline
 & RMSE & MAE & Coverage & PI Width & Moran's I & Shapiro-Wilk\\ 
  \hline
GDEF & 0.410 & 0.331 & 0.904 & 1.450 & $p=0.460$ & $p=0.262$\\ 
  Mat\'{e}rn & 0.436 & 0.355 & 0.936 & 1.689 & $p=0.000$ & $p=0.028$\\ 
  H\&H & 0.684 & 0.552 & 0.898 & 2.209 & $p=0.000$ & $p=0.934$\\ 
  Std. CAR & 0.421 & 0.340 & 0.940 & 1.688 & $p=0.000$ & $p=0.002$\\ 
  CAR 2-param. & 0.422 & 0.342 & 0.936 & 1.647 & $p=0.000$ & $p=0.008$\\ 
   \hline
\end{tabular}
\caption{Comparison of predictive metrics and results from assumption tests across model. In addition to having the smallest out-of-sample prediction error, the GDEF model is the only one that produces estimated residuals that exhibit no spatial correlation.}\label{cvresults}
\end{table}

{The GDEF model has the lowest prediction error among models considered. In addition to this, it is the only model for which the Moran's I statistic results in a failure to reject the null hypothesis of no spatial dependence in the model's residuals. In other words, the GDEF model is the only model among these options with a covariance structure that is capable of fully describing the spatial dependence pattern present in the wheat yield data. Among models compared, the GDEF model produced the narrowest prediction intervals, though the coverage rate of these intervals appears to be somewhat under target, suggesting that the estimated marginal variance of predictions from the GDEF model may be too low; further investigation of this could be warranted. Despite considerably greater model complexity than the alternatives, the GDEF model's out-of-sample predictive performance suggests that it is not overfit to these data and represents an improvement over simpler, more standard methodologies. While it could be argued that the differences in predictive performance between the GDEF and other models is not dramatic, it seems clear that the underlying covariance structure of the wheat yield data is most fully captured by our proposed methodology. We find this to be notable given the dataset's history and reputation as a challenging spatial modeling problem.}

\section{Discussion}

Within this article we introduced a framework for estimating edge weight matrices using the eigenvectors of the line graph's Laplacian matrix. {Utilizing matrix calculations for the derivatives of the log likelihood with respect to basis function coefficients}, we show that this method may be fit efficiently and how it can be used to enhance existing covariance models for areal data. {We also evaluated properties of our approach in a series of simulations and examples that demonstrated its capacity to represent complex covariance structures while significantly improving the computational efficiency of the GDEF model introduced by \citet{christensen2024}.} 
We conclude this article with a brief discussion of potential avenues for future work and model extensions.


{Part of our preference for using the GDEF model in conjunction with the LGL basis rather than the CAR model} stems from findings that the latter characterizes fundamentally non-smooth spatial processes. \citet{rue2005} and \citet{lindgren2011} demonstrate that it is possible to define Markov random fields (MRFs) that approximate smooth spatial processes by allowing some edge weights to be negative. The space of non-CAR MRFs is worth exploring due to their computational advantages. It may be possible to adapt our method for edge weight estimation to allow for the inclusion of negative weights in sparse precision matrix formulations, though additional work would be necessary to ensure that a valid covariance matrix is always produced.

While we have discussed estimation of $\bW$ within the context of explicitly defining a covariance matrix, there are other settings in which a coherent framework for flexibly defining an edge weights matrix may be useful. For example, resistance distance is a graph metric constructed using a matrix of non-negative edge weights that has seen widespread use in many statistical and ecological applications \citep{klein1993,hanks2013,dickson2019}. Outside the context of distances and covariance, an edge weights matrix can be used to characterize the connectivity of different entities within any network.

An additional topic we would like to explore is how our method could be used to characterize the covariance of spatial processes on manifolds or restricted spatial domains. While most spatial models have been developed assuming that data are observed on some well-defined subset of $\mathbb{R}^2$, there is increased interest in the development of methods that can be applied to more complex domains in which Euclidean distances between observations do not accurately represent proximity with respect to the domain's underlying geometry \citep{yang2016}. Mathematical distinctions exist between settings in which data are observed in Euclidean space but concentrated near a lower-dimensional manifold such as the surface of a sphere and settings in which an $n$-dimensional spatial process exists only on an irregular subset of $\mathbb{R}^n$ due to underlying physical constraints \citep{dunson2022}. 
Whether data is observed on a true manifold or a restricted domain, we could define a graph based on the set of observed locations, defining edges between locations based on neighbors within a small radius or through use of the Delauney triangulation \citep{hjelle2006}. The LGL eigenvectors of this graph could then be utilized as in this article to define a valid covariance model over the complex spatial domain; furthermore, the flexibility of our edge weight estimation approach would ultimately allow covariance to be nonstationary with respect to the domain's underlying geometry. 

\section*{Supplemental Materials}

File containing additional simulation results, details on computation, and further discussion of relevant topics. Supplemental materials available online.

\bibliographystyle{apalike}
\bibliography{ref}

\end{document}


\setstretch{1.7}
\maketitle
\vspace{-2em}

The GDEF covariance model used in the main article is defined as follows:
\begin{equation}\label{prevmodel}
\begin{aligned}
    \Sigma_{ij} &= \sigma^2 \rho_\nu(d_{ij}) \\
    d_{ij} &= \sqrt{(\be_i - \be_j)^\top\{\bL^+\}^2(\be_i-\be_j)} \\
    \bL &= \text{diag}(\bW \bone_p) - \bW
\end{aligned}
\end{equation}
where $\sigma^2$ is a scale parameter, $\rho_\nu(\cdot)$ is the Mat\'{e}rn correlation function with smoothness parameter $\nu$ and edge weights matrix $\bW$ is defined according to
\begin{equation}\label{edgebasis}
\begin{aligned}
    \bw &= \text{exp}(\bV^k \eta), \; \eta \in \mathbb{R}^k \\
\end{aligned}
\end{equation}
where $\bV^k$ represents the first $k$ eigenvectors from the LGL basis.

The CAR model used in our work is defined using:
\begin{equation}\label{carmarg2}
    \by \sim N_p(\bzero, \sigma^2 (\text{diag}(\bW \bone_p) - \kappa\bW)^{-1}),
\end{equation}
where $\text{diag}(\cdot)$ returns a diagonal matrix with entries equal to the input vector, with $\kappa$, controlling the degree of correlation between neighboring observations.

\section{Properties of the CAR and GDEF Approaches}

We have stated that our framework for parameterizing the edge weights matrix of a given graph may be used when defining both CAR and GDEF models, but provided little discussion so far regarding the implications of using one covariance model or another. 

CAR models have sparse precision matrices, allowing for computationally efficient likelihood calculations. In contrast, the GDEF model specifies a dense covariance matrix; even when using the proposed basis function parameterization of $\bW$, it is virtually impossible to use the GDEF covariance for graphs with over 10,000 nodes due to the impracticality of inverting matrices of that size. The computational advantages of the CAR approach do however come with trade-offs. 

Several authors have noted that models which condition only on local or nearest-neighbor structure perform more poorly than those incorporating longer range dependencies \citep{gramacy2015,guinness2018,katzfuss2021}. Realizations of spatial random fields produced by CAR models (including ICAR models, which induce perfect correlation between neighboring regions) are considerably less smooth than those generated by the GDEF model, a contrast that is more noticeable on graphs with a large number of nodes. \citet{besag2005} show that the standard ICAR model on a regular two-dimensional lattice grid converges asymptotically to the distribution of a two-dimensional Brownian motion as grid resolution increases, while \citet{lindgren2011} show that the ICAR model may be thought of as a discrete approximation to the limit of a Mat\'{e}rn random field whose range parameter goes to infinity and smoothness goes to zero. \citet{paciorek2013} demonstrates that CAR models using more than first order adjacency (extending graph structure to second order neighbors and beyond) do not result in meaningfully smoother fields. While computationally efficient, straightforward to implement and widely used, CAR models carry implicit assumptions that may not be desirable in practice.

In contrast to the CAR model, the GDEF model accommodates a wider range of spatial dependence patterns. Equation \ref{prevmodel} defines the GDEF covariance, with Mat\'{e}rn correlation function: 

\begin{equation}\label{our_mat}
    \rho_\nu(d) = \frac{2^{1-\nu}}{\Gamma(\nu)}\left( \sqrt{2\nu}d\right)^\nu K_\nu \left( \sqrt{2\nu}d\right).
\end{equation}
Here $d$ refers to distance, $\nu > 0$ is a smoothness parameter, often fixed in practice, and $K_\nu$ is a modified Bessel function of the second kind of order $\nu$. The exponential covariance function $\sigma^2 \text{exp}(-d)$ is a special case of the Mat\'{e}rn covariance when $\nu=1/2$; $\nu = 3/2$ and $\nu = 5/2$ are popular choices for spatial models, and the squared exponential covariance $\sigma^2 \text{exp}(-d^2/2)$ is the limit of Equation \ref{our_mat} as $\nu$ goes to infinity. Those familiar with the Mat\'{e}rn family of covariance functions may note that Equation \ref{our_mat} omits an explicit range parameter. As discussed in \citet{christensen2024}, the range parameter of the GDEF model is implicitly defined by the scale of the edge weights matrix, e.g. multiplying $\bW$ by a factor of two is equivalent to doubling the range parameter. 

\begin{figure}
    \centering
    \includegraphics[width = .9\linewidth]{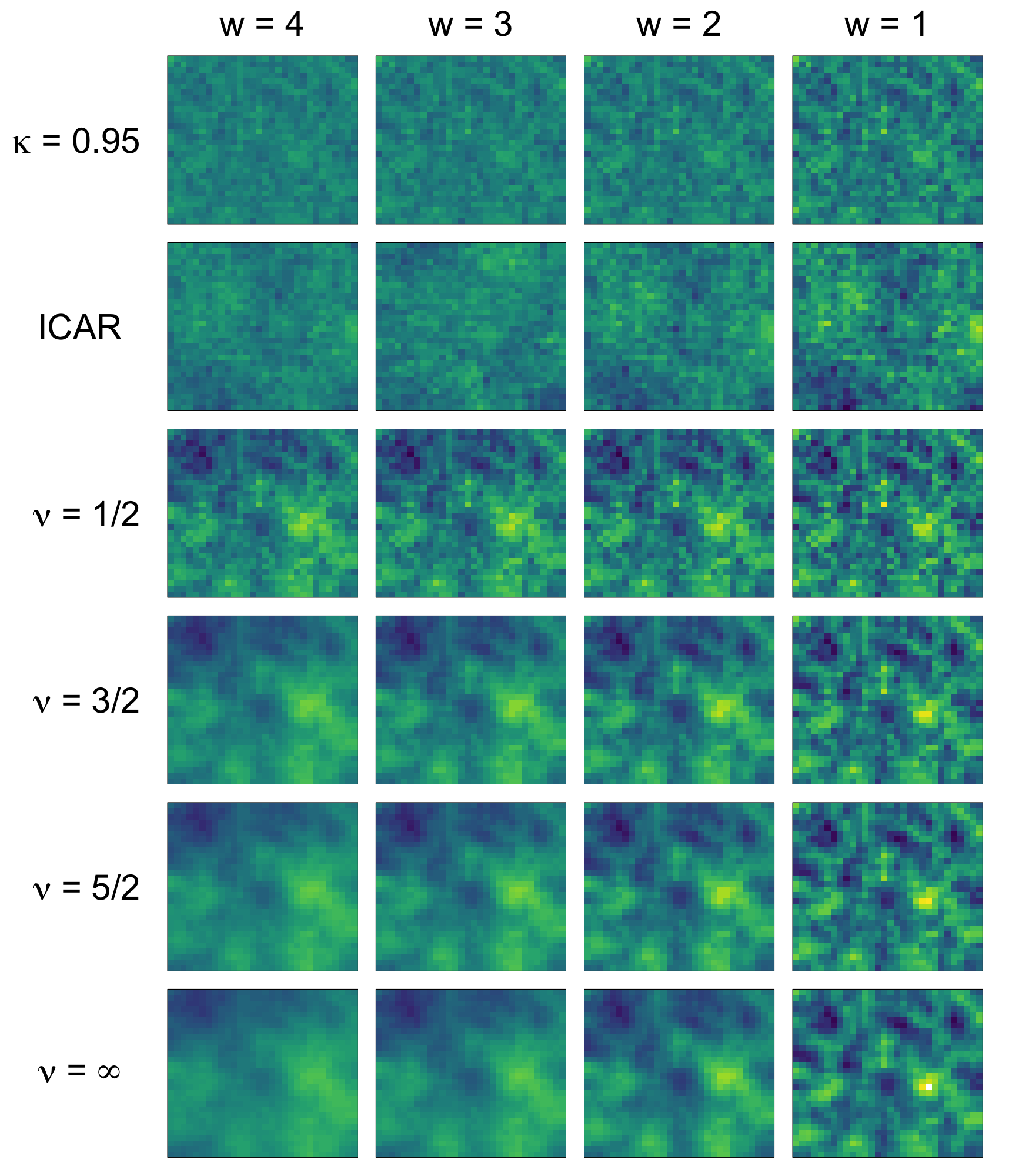}
    \caption{Samples from $N(\bzero,\Sigma)$ over a $30 \times 30$ grid with uniform edge weights. Columns correspond to the weight of all nonzero entries in $\bW$, while rows correspond to a different parameterizations of $\Sigma$. The first row uses the CAR covariance with $\kappa = 0.95$, the second uses the ICAR covariance, and subsequent rows use the GDEF covariance for different values of $\nu$. Note that all samples were produced using the same seed.}
    \label{covcomp}
    \end{figure}

Figure \ref{covcomp} highlights some distinctions between the CAR and GDEF approaches. Depicted samples were generated over a $30 \times 30$ grid from a mean zero normal distribution. Each row corresponds to a different covariance model: the CAR model from Equation \ref{carmarg2} with $\kappa = 0.95$, the ICAR model with a sum-to-zero constraint (the ``smoothest" possible CAR model) and the GDEF model with four different values of $\nu$. In the first column, $\bW$ was constructed such that $w_{ij} = 4$ for all $i\sim j$, with different constants used in subsequent columns. Across all settings the scale parameter $\sigma^2$ was set equal to 1. As expected, samples produced by the CAR models and the GDEF model for $\nu=1/2$ are visibly less smooth than those from GDEF models with larger values of $\nu$. The impact that the scale of $\bW$ has on each model is also notable; samples with edge weights equal to 4 exhibit slower decay in spatial correlation than those with smaller edge weights. For CAR models, the scale of $\bW$ does not influence range, but rather the scale of marginal variances. 

In comparing the CAR and GDEF approaches, there has been discussion of smoothness and it would be easy to think about in terms of whether or not this is inherently a ``good" or ``bad" thing. ``More smooth = good" is clearly a misguided framing; one of the common criticisms of basis function based approaches is a tendancy to oversmooth, something that our approach avoids by embedding the basis functions in the covariance rather than defining spatial random effects as linear combinations of eigenvectors. Because the GDEF framework allows for the use of any covariance function that is valid for Euclidean distances, one can calibrate the smoothness level as desired to the application in question. When using the CAR, one is forced into specifiying a model where spatial dependence exhibits random walk/Brownian motion-like behavior. This may be valid in some circumstances but certainly not all. 

This gets to perhaps a more important point regarding our preference for the GDEF model over the CAR model. In many ways, thinking about CARs as covariance models rather than in terms of their conditional specifications probably does them a disservice. \citet{wall2004close} and others have noted that there are some fairly concerning issues with the covariance structure implied by CARs, demonstrating that pairwise correlations can change dramatically when one edge or node is changed. I think they are quite useful for defining/estimating spatial random effects, or as a prior for such effects, but using them as a generative model often leads to less desirable behavior. Because so much of our work with edge weight estimation is focused specifically on the question of inference on the covariance structure, we feel it is natural to prefer the more ``well-behaved" GDEF model, which leans on the well studied behavior of the Mat\'{e}rn covariance to the CAR, which wasn't really developed to describe covariance structures in the first place. 

\section{CAR Models: range and resolution}
\begin{figure}
    \centering
    \includegraphics[width = .9\textwidth]{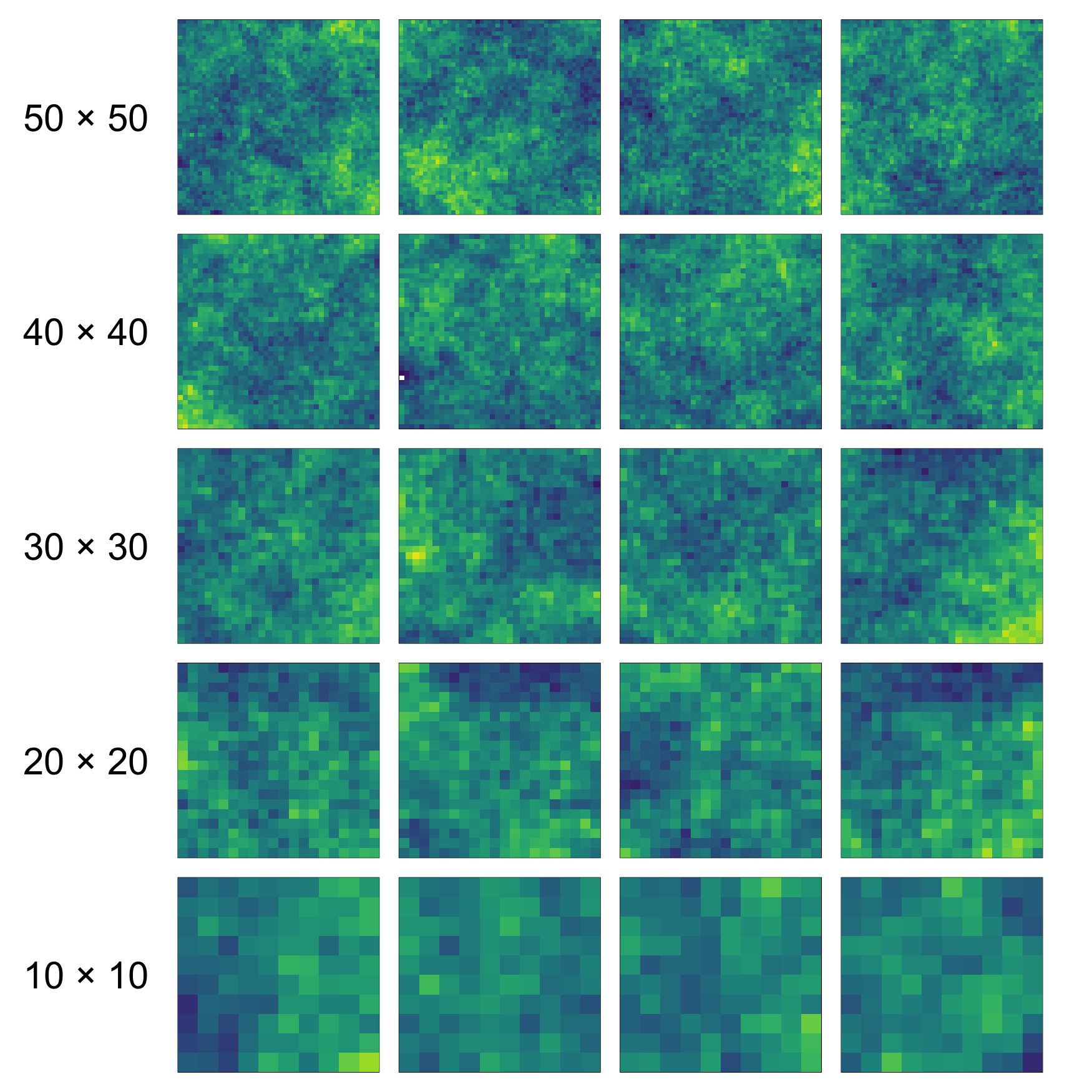}
    \caption{Samples from an ICAR model with uniform edge weights at different grid resolutions.}
    \label{icarres}
\end{figure}

The notion of range within CAR models is trickier to characterize than for GDEF models. While the correlation parameter $\kappa$ ostensibly controls the smoothness of the Markov random field and the rate of spatial decay between areal regions, it is highly sensitive to the resolution of the spatial domain. This may be particularly relevant when the CAR model is viewed as the discretization of some continuous spatial process. As areal resolution becomes finer, the space of region-averaged continuous spatial processes that could be represented using a CAR or ICAR model shrinks. Figure \ref{icarres} highlights this, depicting samples from an ICAR model at different spatial resolutions. If each grid represents the same subset of $\mathbb{R}^2$ and samples are viewed as the region-averaged realizations of continuous spatial processes, samples over coarser grids could correspond to smoother underlying processes than those generated at higher resolutions.

\section{Incorporating covariate information}

We may be interested in the question of whether certain environmental features inhibit or facilitate connectivity between regions in our spatial domain. As such we may wish to model edge weights as a function of environmental covariates.  Generally speaking, we will not have covariates that are explictly associated with the edges of our graph. If $\bX$ is a $p \times r$ matrix containing $r$ features for each of the $p$ nodes in a graph, \citet{hanks2013} suggested defining $\bX^E$, a $q \times r$ feature matrix for the graph's edges, by averaging the features of adjacent nodes:
\begin{equation}
    \bx^{e_{ij}} = \frac{\bx_i + \bx_j}{2}.
\end{equation}
Here $\bx^{e_{ij}}$ is the row of $\bX^E$ corresponding to the edge linking nodes $i$ and $j$. The matrix $\bX^E$ could also be constructed treating differences in the features of adjacent regions as a covariate itself. Others have suggested that the length of the shared border between two adjacent regions may be important to consider. However $\bX^E$ is defined, we can augment the basis function model for edge weights in Equation \ref{edgebasis} by setting 
\begin{equation}\label{envmodel}
    \bw = \text{exp}(\bX^E\psi + \bV^k\eta), \; \psi \in \mathbb{R}^r, \; \eta \in \mathbb{R}^k.
\end{equation}

In order to avoid confounding between $\psi$ and $\eta$ due to correlation between $\bX^E$ and $\bV^k$, let $\bV^\perp$ contain the eigenvectors of $\bX^\perp \bL^L \bX^\perp$, where $\bX^\perp = (\bI_q - \bX^E(\bX^{E\top}\bX^E)^{-1}\bX^{E\top})$ is the projection onto the orthogonal compliment of the column space of $\bX^E$. The $q \times k$ matrix $\bV^{k\perp}$ contains the $k$ eigenvectors corresponding to the smallest eigenvalues of$\bX^\perp \bL^L \bX^\perp$. The matrix $\bX^E$ is orthogonal to $\bV^{k\perp}$, which may improve estimation and interpretation of $\psi$. This approach to spatial deconfounding was introduced by \citet{reich2006} and has seen widespread use since then, though there is debate as to whether such adjustments are appropriate \citep{zimmerman2022}. While our goal is not to contribute to this discussion from a theoretical perspective, we find that using the orthogonalized $\bV^{k \perp}$ results in a slightly more flexible covariance model when weights are defined using Equation \ref{envmodel}.

\section{Replicate observations}
We note in the main manuscript that the spatial deformation literature traditionally assumes the setting in which multiple observations of the spatial process is made. This is because many implementations of the method utilize the empirical covariance of the replicates as the primary input within their models, see for example \citet{sampson1992,schmidt2003}. 

A natural question to ask is whether not it is realistic to assume that multiple independent observations of some spatial process will commonly exist in practice. Generally when such methods have been utilized, the replicate observations are actually spatiotemporal observations, where repeated measurements have been made over time at a collection of spatial locations. In order to implement the spatial deformation approach, one must either assume that the observations are temporally independent, or alternatively, one can define a separable space-time covariance structure.

One of the advantages of our proposed methodology is that the integration of our dimension reduction approach to edge weight estimation within the GDEF covariance model produces a framework that allows us to fit models with no replicate observations of the spatial process that provide an interpretation similar to the spatial deformation approach for modeling nonstationarity.

\section{Derivatives of model likelihood}

Below we provide details on how to obtain $\nabla_\ell(\theta)$ the partial derivative of the log-likelihood $\ell$ with respect to an individual basis coefficient $\eta_i$; we assume here that $\bY \sim N(\bzero, \bSigma\otimes \bI_n)$ and $\Sigma$ is defined by the GDEF covariance with $\nu = 3/2$. One may obtain the derivative $\frac{\partial\ell}{\partial \eta_i}$ through repeated application of the chain rule as follows:

\begin{eqnarray}
\frac{\partial\ell}{\partial \eta_i} &=& -\frac{n}{2} \mbox{trace} \left( \bSigma^{-1} \frac{\partial\bSigma}{\partial \eta_i} \right) + \frac{1}{2} \sum_{r=1}^n \by_r' \bSigma^{-1} \frac{\partial\bSigma}{\partial \eta_i} \bSigma^{-1} \by_r \nonumber \\
\frac{\partial\bSigma}{\partial \eta_i} &=& -\sigma^2\frac{\partial\bD}{\partial \eta_i} \odot \bD \odot \exp(-\bD)= -\frac{\sigma^2}{2}\frac{\partial\bD_2}{\partial \eta_i} \odot \exp(-\bD), \nonumber  \\
\frac{\partial\bD_2}{\partial \eta_i} &=& \mbox{diag} \left( \frac{\partial\{\bL^+\}^2}{\partial \eta_i} \right) \mathbf{1}'_n + \mathbf{1}_n \mbox{diag} \left( \frac{\partial\{\bL^+\}^2}{\partial \eta_i} \right)'-2\frac{\partial\{\bL^+\}^2}{\partial \eta_i},  \\
\frac{\partial\{\bL^+\}^2}{\partial \eta_i} &=& \bL^+ \frac{\partial\bL^+}{\partial \eta_i} +\frac{\partial\bL^+}{\partial \eta_i} \bL^+, \nonumber\\
\frac{\partial\bL^+}{\partial \eta_i} &=& -\bL^+ \frac{\partial \bL}{\partial \eta_i} \bL^+, \nonumber  \\
\frac{\partial \bL}{\partial \eta_i} &=& \mbox{diag} \left( \frac{\partial\bW}{\partial \eta_i} \mathbf{1}_n \right)-\frac{\partial\bW}{\partial \eta_i} \nonumber. 
\end{eqnarray}

Here $\bD_2 = \bD \odot \bD$ and $\frac{\partial\bW}{\partial \eta_i}$ is a $p \times p$ matrix equal to $\bW$ with each of its nonzero elements multiplied by the corresponding element of basis vector $\bv_i$. In order to implement Newton's method, one also needs $\bH_\ell(\eta)$, the second derivative of the log likelihood with respect to $\eta$. The second derivative of the GDEF covariance with respect to $\eta$ is given as: 
\begin{eqnarray}
\frac{\partial^2\bSigma}{\partial\eta_i \partial\eta_j} &=& -\sigma^2\exp(-\bD) \odot \left[\frac{d^2\bD}{\partial\eta_i\partial\eta_j} \odot \bD+\frac{\partial\bD}{\partial\eta_i} \odot \frac{\partial\bD}{\partial\eta_j}-\frac{\partial\bD}{\partial\eta_i} \odot \frac{\partial\bD}{\partial\eta_j} \odot \bD\right], \nonumber  \\
 &=& -\frac{\sigma^2}{2}\exp(-\bD) \odot \left[\frac{d^2\bD_2}{\partial\eta_i\partial\eta_j} -\frac{\partial\bD_2}{\partial\eta_i} \odot \frac{\partial\bD_2}{\partial\eta_j}\oslash 2\bD \right] \text{ (Let 0/0 = 0)}, \nonumber  \\
\frac{\partial\bD_2}{\partial\eta_i} &=& 2\frac{\partial\bD}{\partial\eta_i} \odot \bD, \hspace{3mm} i=1,\ldots,k \nonumber \\
\frac{d^2\bD_2}{\partial\eta_i\partial\eta_j} &=& \mbox{diag} \left( \frac{d^2\{\bL^+\}^2}{\partial\eta_i\partial\eta_j} \right) \mathbf{1}'_n + \mathbf{1}_n \mbox{diag} \left( \frac{d^2\{\bL^+\}^2}{\partial\eta_i\partial\eta_j} \right)'-2\frac{d^2\{\bL^+\}^2}{\partial\eta_i\partial\eta_j},  \\
\frac{d^2\{\bL^+\}^2}{\partial\eta_i\partial\eta_j} &=& \bL^+ \frac{d^2\bL^+}{\partial\eta_i\partial\eta_j}+\frac{\partial\bL^+}{\partial\eta_j}\frac{\partial\bL^+}{\partial\eta_i} +\frac{d^2\bL^+}{\partial\eta_i\partial\eta_j} \bL^+ +\frac{\partial\bL^+}{\partial\eta_i}\frac{\partial\bL^+}{\partial\eta_j}, \nonumber \\
\frac{d^2\bL^+}{\partial\eta_i\partial\eta_j} &=& [\bL^+ \frac{\partial\bL}{\partial\eta_j} \bL^+]\frac{\partial\bL}{\partial\eta_i}\bL^+-\bL^+ \frac{d^2\bL}{\partial\eta_i\partial\eta_j}\bL^+ +\bL^+\frac{\partial\bL}{\partial\eta_i}[\bL^+ \frac{\partial\bL}{\partial\eta_j} \bL^+], \nonumber  \\
\frac{d^2\bL}{\partial\eta_i\partial\eta_j} &=& \mbox{diag} \left( \frac{d^2\bW}{\partial\eta_i\partial\eta_j} \mathbf{1}_n \right)-\frac{d^2\bW}{\partial\eta_i\partial\eta_j}. \nonumber
\end{eqnarray}
Here, the second derivatives of matrix $\bW$ with respect to $\eta$ have entries equal to $\bw \odot \bv_i \odot \bv_j$.

The second derivative with respect to the likelihood is given by
\begin{eqnarray}\label{eq:derloglik2}
 \frac{d^2\ell}{\partial\eta_i\partial\eta_j} &=& -\frac{n}{2} \mbox{trace} \left( \bSigma^{-1} \frac{d^2\bSigma}{\partial\eta_i\partial\eta_j}- \bSigma^{-1} \frac{\partial\bSigma}{\partial\eta_i} \bSigma^{-1} \frac{\partial\bSigma}{\partial\eta_j}\right) \nonumber \\
 &-& \sum_{r=1}^n \by_r' \bSigma^{-1} \frac{\partial\bSigma}{\partial\eta_i} \bSigma^{-1} \frac{\partial\bSigma}{\partial\eta_j} \bSigma^{-1} \by_r+ \sum_{r=1}^R \frac{1}{2} \by_r'\bSigma^{-1} \frac{d^2\bSigma}{\partial\eta_i\partial\eta_j} \bSigma^{-1} \by_r.
\end{eqnarray}

We have that $E(\by_r'\bA\by_r)=\mbox{trace}(\bA\bSigma)$.
The expected value of this second derivative of the likelihood, with respect to the data $\bY$ is
\begin{equation}\label{eq:Ederloglik2}
E \left[ \frac{d^2\ell}{\partial\eta_i\partial\eta_j} \right]= -\frac{n}{2} \mbox{trace} \left( \bSigma^{-1} \frac{\partial\bSigma}{\partial\eta_i}\bSigma^{-1} \frac{\partial\bSigma}{\partial\eta_j} \right).
\end{equation}
The expected second derivative of the log likelihood can be computed considerably faster than $\bH_\ell(\eta)$, leading us to recommend its use during the iterative portion of the implementation of Newton's method, and computing the full second derivative only once after a maximum has been obtained.

While other optimization schemes, such as those found in \citet{nocedal2006numerical}, could potentially be used to obtain the MLE of $\theta$, chapter 14 of \citet{lange2010numerical} suggests that Fisher scoring will work well in settings such as ours where we have a close form expression for the expected Hessian. In practice we have found this to be the case, though convergence can be slow depending on the size of the graph and number of basis functions being used.

\section{Additional Simulation and Details From Main Text Studies}

\subsection*{Simulation S1: Approximate confidence interval coverage}

We defined $10 \times 10$ and $20 \times 20$ lattices, and obtained the LGL eigenvectors for each. We selected $k$ eigenvectors and simulated $\eta \sim N_k(\bzero,\Phi)$, where $\Phi$ is a diagonal matrix with first entry (corresponding to the variance of the ``intercept" eigenvector) equal to 0.5 and all other diagonal entries equal to 25. We then generated $n$ samples from $N_p(\bzero,\Sigma(\eta))$ where $\Sigma(\eta)$ is defined by Equations \ref{prevmodel} and \ref{edgebasis}, using a Mat\'{e}rn covariance with $\nu = 3/2$ and fixing $\sigma^2=1$. This was done for each of $k \in \{10,30\}$ and $n \in \{1,10,50\}$. We obtained the MLE $\hat{\eta}$ and calculated approximate 90\% confidence intervals for $\eta$ using Fisher's approximation. The simulation was repeated ten times, with average coverage rates across all elements of $\eta$ provided in Table \ref{sim1}.

\begin{table}[ht]
\centering
\begin{tabular}{r|ccc}
  \hline
 & $n=1$ & $n=10$ & $n=50$ \\ 
  \hline
$p=100, \; k= 10$ & 0.880 ($\pm$ 0.039) & 0.910 ($\pm$ 0.023)& 0.870 ($\pm$ 0.040) \\ 
  $p=100, \; k= 30$ & 0.804 ($\pm$ 0.040)& 0.867 ($\pm$ 0.027)& 0.927 ($\pm$ 0.015)\\ 
  $p=400, \; k= 10$ & 0.920 ($\pm$ 0.025)& 0.910 ($\pm$ 0.023)& 0.910 ($\pm$ 0.031)\\ 
  $p=400, \; k= 30$ & 0.907 ($\pm$ 0.018)& 0.900 ($\pm$ 0.021)& 0.883 ($\pm$ 0.019)\\ 
   \hline
\end{tabular}
\caption{Coverage rates of approximate 90\% confidence intervals for $\eta$. Monte Carlo standard error in parentheses.}\label{sim1}
\end{table}

In general coverage rates are appropriate, the exception being under the settings $k =30$ and $n=1$ over a $10 \times 10$ lattice. On top of having lower than expected coverage, Newton's method diverged in one of the ten simulations under these settings, indicating that one observation of a spatial process over a graph with $p=100$ nodes may not be enough to inform the estimation of $\Sigma$ using $k=30$ basis functions. In general however, we feel that inference based on Fisher's approximation to the distribution of the MLE is appropriate. We investigate choice of $k$ in the following simulation.

\subsection*{Additional detail from Main Section 4.1: Number of basis functions}

\begin{figure}
    \centering
    \includegraphics[width = .95\textwidth]{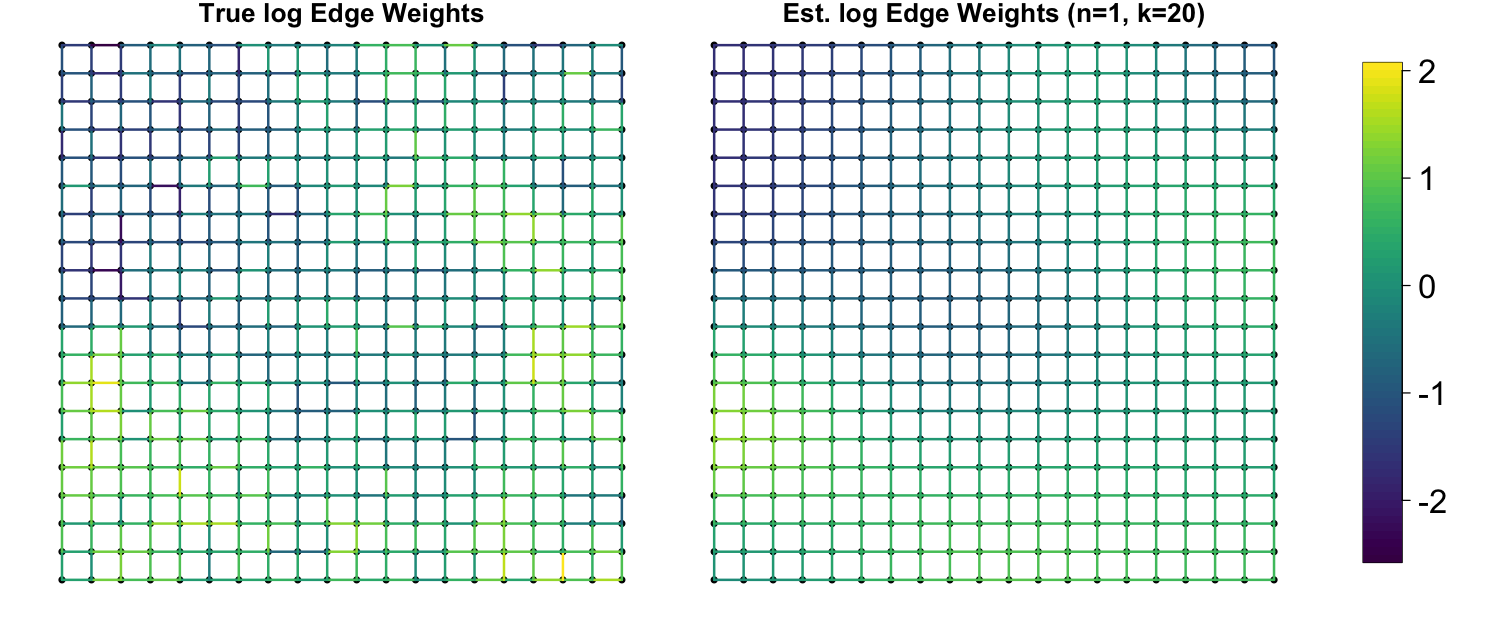}
    \caption{Referenced in main text. Simulated $\bW$ on left, with $\hat{\bW}$ on right (estimated from $n=1$ observations and using $k=20$ basis functions). Approximation ``smooths" the true edge weights.}
    \label{icarweights}
\end{figure}

Within this study we compared AIC and BIC across different model fits. AIC and BIC are considered not because they are objectively the best tools for model selection, but because they provide a likelihood-based starting point for the question of how to choose the number of eigenvectors when implementing our method. BIC is more conservative than AIC due to a larger model size penalty. The two criteria also differ in that AIC is calibrated for improving predictive accuracy, while BIC may be viewed as favoring the model that best approximates the marginal distribution of the data \citep{gelman2014}. The fact that the two criteria frequently prefer different values of $k$ provides some intuition for the range on model size that may be appropriate, which led us to the suggestion that a reasonable starting point is to select $k=\sqrt{np}$ when the model reasonably approximates the true data generating process, as was the case within this simulation. 

Table \ref{diverge} contains the rates at which the Fisher scoring algorithm used to estimate $\hat{\eta}$ failed to converge across different simulation settings, with the majority of instances occurring for large $k$ and $n=1$. When computational instability is encountered, refitting with smaller $k$ will frequently resolve the issue; the possibility of instability due to significant model misspecification should however be considered.

\begin{table}[ht]
\centering
\begin{tabular}{l|rrrrrrr}
  \hline
 & $k=40$ & $k=50$ & $k=60$ & $k=70$ & $k=80$ & $k=90$ & $k=100$ \\ 
  \hline
$p=100, \; n=1$  & 0.3 & 0.7 & 0.8 & 1.0 & 1.0 & 1.0 & 1.0 \\ 
  $p=100, \; n=10$ & 0.0 & 0.0 & 0.0 & 0.0 & 0.0 & 0.2 & 0.1 \\ 
  $p=100, \; n=50$  & 0.0 & 0.0 & 0.0 & 0.0 & 0.0 & 0.0 & 0.0 \\ 
   \hline
\end{tabular}
\caption{Frequency with which Fisher scoring algorithm failed to converge. Estimation was stable for all $k < 40$ and for all settings over the $20 \times 20$ lattice.}\label{diverge}
\end{table}
\subsection*{Additional detail from Main Section 4.2: Model performance under misspec.}

\subsubsection*{Generating random deformations}
In order to obtain random deformations of the initial $15 \times 15$ grid, we utilize the following steps. Let $s_i$ be the location of the $i$th grid point.

\begin{figure}
    \centering
    \includegraphics[width=0.6\linewidth]{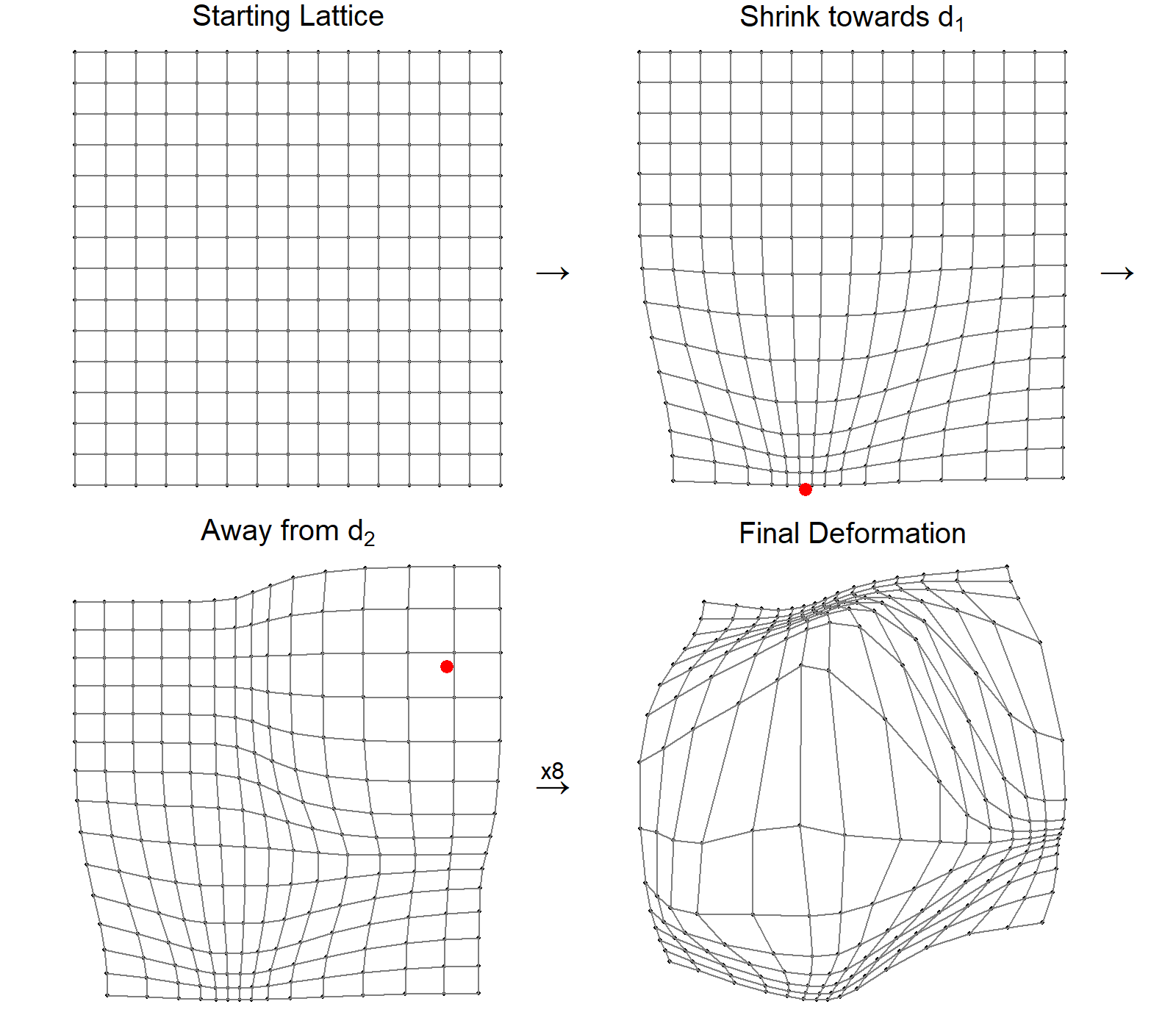}
    \caption{Illustration of how random deformations are generated}
    \label{deform}
\end{figure}

\begin{itemize}
    \item Randomly sample a new location $d^*$ within the boundaries of the grid
    \item For $i \in 1:225$ obtain the vectors $v_i = s_i - d^*$
    \item For each vector $v_i$, compute $c_i = 3/5 *exp(-||v_i||_3/50)$
    \item If taking an expansion step, update the location of all points by defining $s_i = s_i + c_i v_i$, if taking a contraction step, update them using $s_i = s_i - c_i v_i$
    \item For 10 iterations, alternate between expansion and contraction steps
\end{itemize}
This process is visualized in Figure \ref{deform}. While the development of this approach for defining random deformations was somewhat ad hoc, we found that it worked well for the purposes of this simulation; it is also worth noting that the design of the above algorithm was guided by the need to ensure that the deformation did not result in a folding of the space onto itself, which can lead to invalid covariance structures, as discussed by \citet{sampson1992}.

\subsubsection*{Implementing the S\&G model using the \texttt{deform} package}

To implement the model of \citet{sampson1992} within our simulation study we utilized the \texttt{deform} package in R \citep{deformR}. The package author notes that their implementation of the S\&G model in its current iteration (version 1.0.0) is fairly slow, and we found this to be the case. The default settings of the package are to use a 10 by 10 grid of knots when fitting the thin plate splines used to estimate the spatial deformation. We found that this led to computational instability and long compute times (upwards of an hour) when fit using limited numbers of replicates. For $n=2$, $n=3$, and $n=5$ we found that KL divergences were best when using a 5 by 5 grid of knots, and compute times were markedly faster (roughly 4-5 minutes). Larger number of replicates supported the use of more knots, leading to better estimation of the deformation. We used a 7 by 7 grid for $n=10$, a 10 by 10 grid for $n=25$, and a 12 by 12 grid for $n=50$, with the latter settings generally requiring 8-10 minutes of compute time. For reference, the GDEF model using 15 basis functions required 30-40 seconds to run, and the \texttt{geoR} package required $<10$ seconds to estimate the Mat\'{e}rn covariance parameters on the 15 by 15 grid of locations.

\section{Code and Data}
The code used to conduct the analyses discussed within this manuscript is available at the JCGS website. The wheat yield data set from \citet{mercer1911} is contained in the spData R package. It can be accessed by loading the package and running the command ``data(wheat)".

\bibliographystyle{apalike}
\bibliography{ref}